\documentclass[twocolumn,tighten]{aastex62}
\usepackage{amsmath,bm,graphicx,color} 

\newcommand{\dtime}{\partial_t}
\newcommand{\vcz}{\mathrm{v}_z}
\newcommand{\zhat}{\boldsymbol{\hat{z}}}
\newcommand{\vvb}{\boldsymbol{\mathrm{v}}}
\newcommand{\vkb}{\boldsymbol{k}}
\newcommand{\vrb}{\boldsymbol{r}}
\newcommand{\bomega}{\boldsymbol{\Omega}}
\newcommand{\grad}{\boldsymbol{\nabla}}
\newcommand{\dvg}{\boldsymbol{\nabla}\boldsymbol{\cdot}}

\newcommand{\cnabla}{\!\boldsymbol{\cdot}\!\boldsymbol{\nabla}}
\newcommand{\cross}{\boldsymbol{\times}}
\newcommand{\bcdot}{\boldsymbol{\cdot}}
\newcommand{\bilinear}[2]{{#1}\!\boldsymbol{\cdot}\!{#2}}
\newcommand{\Ro}{\mathrm{Ro}}
\newcommand{\rmv}{\mathrm{v}}

\newcommand{\Roc}{\mathrm{Ro_c}}

\shorttitle{Rotating Convection \& Penetration}
\shortauthors{Augustson et al.}
\begin{document}

\title{A model of rotating convection in stellar and planetary interiors: \\ I - convective penetration} 

\correspondingauthor{K.~C. Augustson} 
\email{kyle.augustson@cea.fr}
\author{K.~C. Augustson}
\author{S. Mathis}

\affil{AIM, CEA, CNRS, Universit\'{e} Paris-Saclay, Universit\'{e} Paris Diderot, Sarbonne Paris Cit\'{e}, F-91191 Gif-sur-Yvette Cedex, France}\label{inst1}

\begin{abstract}
  A monomodal model for stellar and planetary convection is derived for the magnitude of the rms velocity, degree of
  superadiabaticity, and characteristic length scale as a function of rotation rate as well as with thermal and viscous
  diffusivities. The convection model is used as a boundary condition for a linearization of the equations of motion in
  the transition region between convectively unstable and stably-stratified regions, yielding the depth to which
  convection penetrates into the stable region and establishing a relationship between that depth and the local
  convective Rossby number, diffusivity, and pressure scale height of those flows.  Upward and downward penetrative
  convection have a similar scaling with rotation rate and diffusivities, but they depend differently upon the pressure
  scale height due to the differing energetic processes occurring in convective cores of early-type stars versus
  convective envelopes of late-type stars.
\end{abstract}


\defcitealias{stevenson79}{S79}
\defcitealias{zahn91}{Z91}

\section{Introduction}\label{sec:intro}

In the context of stellar and planetary physics, convection driven through buoyancy and doubly-diffusive instabilities
plays a critical role in mixing and transport processes in convectively unstable regions of stars
\citep[e.g.,][]{miesch09,kupka17,garaud18}.  It primarily serves to transport the energy released deep within the star
or planet through regions where radiative energy transport is inefficient.  Moreover, convection directly and nonlocally
transports heat and chemicals through advection and it acts diffusively through entrainment and dissipative processes.
In the presence of rotation, convection can also transport angular momentum through Reynolds stresses and meridional
flows to establish and maintain a differential rotation
\citep[e.g.,][]{glatzmaier82,kichatinov86,kitchatinov93,brummell96,busse02,brun02,miesch09,augustson12,brun17a}. However,
parameterizing the impact of rotation on the convection and its related transport properties over evolutionary time
scales remains a relatively open problem.

Convective flows cause mixing not only in regions of superadiabatic temperature gradients but in neighboring
subadiabatic regions as well, as motions from the convective region contain sufficient inertia to extend into those
regions before being buoyantly braked or turbulently eroded
\citep[e.g.,][]{massager90,hurlburt94,miesch05,lecoanet15,viallet15}.  This convective penetration and turbulence can
thus alter the chemical composition and thermodynamic properties in those regions, softening the transition between
convectively stable and unstable regions, with the consequence being that the differential rotation, opacity, and
thermodynamic gradients are modified
\citep[e.g.,][]{spiegel63,zahn91,canuto92,freytag96,brummell02,browning04,augustson12,brun17a,pratt17b}. Such processes
have an asteroseismic signature as has been observed in many kinds of stars
\citep[e.g.,][]{aerts03,degroote10,neiner12a,briquet12,zhang13,moravveji16,pedersen18}. Indeed, penetrating convection
can modify the depth of the convection zone and any extant tachocline leading to a frequency shift in asteroseismically
detected modes \citep[e.g.,][]{dziembowski91,monteiro00,christensen11,montalban13} and advect magnetic field and angular
momentum into the stable region leading to changes in dynamo action and the differential rotation
\citep[e.g.,][]{miesch05,browning06,browning07,jones10,featherstone09,masada13,augustson13,augustson16}. In stars with a
convective core, convective overshoot and penetration can lead to a greater amount of time spent on stable burning
phases as fresh fuel is mixed into the burning region
\citep[e.g.,][]{mowlavi94,meakin07,arnett09,maeder09,viallet13,jin15}.  Penetrative convection induces chemical mixing
near the transition between convectively stable and unstable regions of stars, leading to changes in the spectral
characteristics of the atmosphere \citep[e.g.,][]{schatzman77,vauclair78,freytag10,baraffe17}. Hence, from the
standpoint of stellar evolution, an open problem is to understand how the depth of penetration and the character of
the convection in this region change with rotation, magnetism, and diffusion.

In this work, a generalized version of a heuristic model for convection for rotating systems is developed following
\citet{stevenson79}. This model of convection is then employed to estimate the convective depth of penetration above and
below a convection zone. The linearized Boussinesq equations of motion that yield the growth rate used in the model are
given in \S\ref{sec:boussinesq}.  The heat-flux maximization principle is discussed in \S\ref{sec:mheat}. The model of
convection is derived for a diffusion-free setting in \S\ref{sec:dfree}, when including thermal diffusion in
\S\ref{sec:thermaldiff}, and when taking into account both thermal and viscous diffusion in \S\ref{sec:viscdiff}.  This
convection model is then used in conjunction with the linearized model of convective penetration derived in
\citet{zahn91} to estimate the depth of convective penetration as shown in \S\ref{sec:penetration}.  A summary of the
results and perspectives are presented in \S\ref{sec:final}.

\section{Heat-Flux Maximized Convection Model}\label{sec:hfmcm}

As a first step, the heuristic model will be considered to be local such that the length scales of the flow are much
smaller than either density or pressure scale heights. This is equivalent to ignoring the global dynamics and assuming
that the convection can be approximated as local at each radius and colatitude in a star or planet.  As such, one may
consider the dynamics to be in the Boussinesq limit (Section \ref{sec:boussinesq}).  The linearized dynamics is used to
construct the heat-flux maximization principle.

\subsection{Linearized Boussinesq Dynamics} \label{sec:boussinesq}

\begin{figure}[t!]
  \begin{center}
    \includegraphics[width=0.5\textwidth]{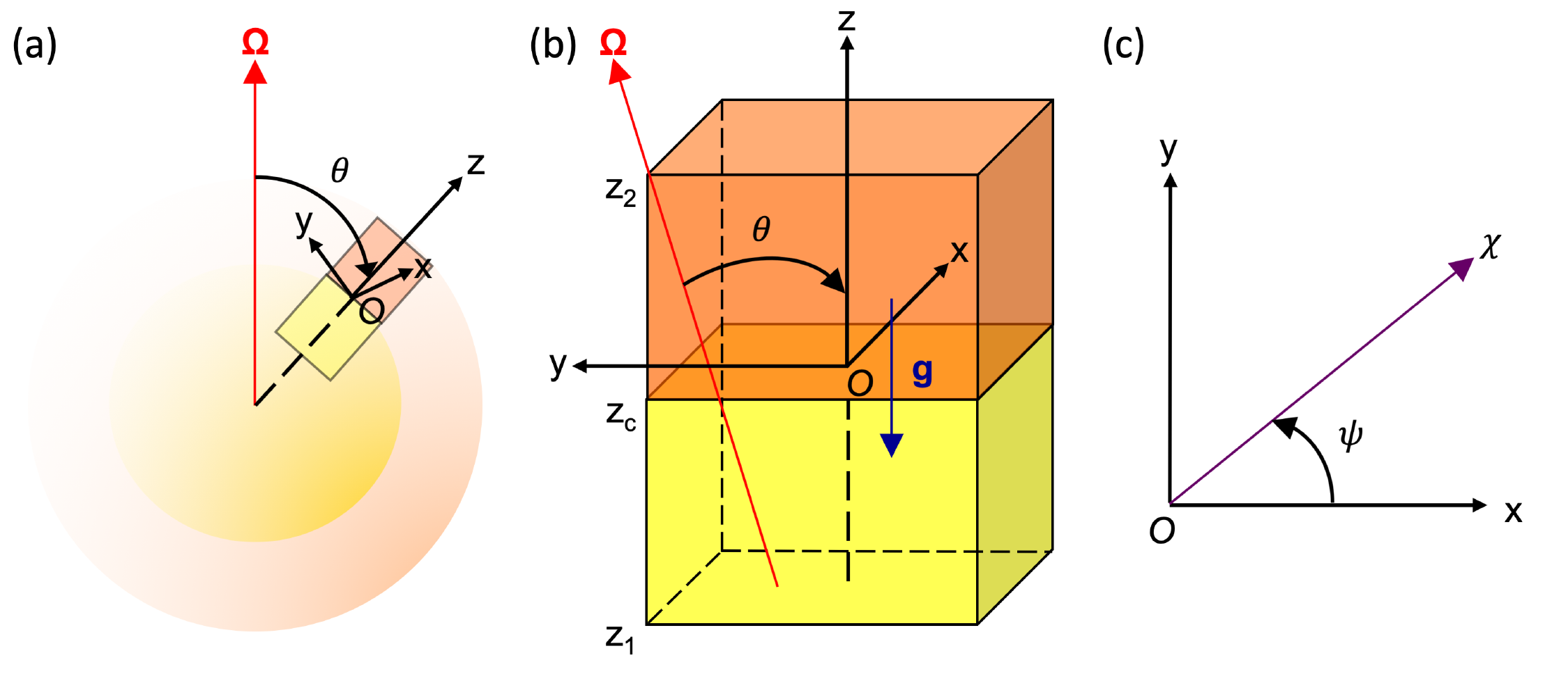} 
    \caption{Coordinate system adopted for the models of rotating convection and penetrative convection, showing (a) the
      global geometry and f-plane localization, (b) the f-plane geometry, and (c) the direction $\chi$ in the horizontal
      plane of the f-plane.  The orange tones denote a convective region and the yellow tones denote a stable region for
      late-type stars, and vice versa in early-type stars. }\label{fig:coords}
  \end{center}
\end{figure}

Boussinesq devised one of the simplest convective systems, consisting of an infinite layer of a nearly incompressible
fluid with a small thermal expansion coefficient $\alpha_T=-\partial\ln{\rho}/\partial T|_P$ that is confined between
two infinite impenetrable plates differing in temperature by $\Delta T=T(z_2)-T(z_c)$, where it is assumed for this model that
$T(z_2)<T(z_c)$, and separated by a distance $\ell_0=z_2-z_c$, as in Figure \ref{fig:coords}. The thermodynamic
variables are further expanded about their averages as $q = \langle q\rangle + \bar{q}(z) + q'$, with $\langle q\rangle$
being the volumetric mean, $\bar{q}$ being the horizontal average, and $q'$ being the dynamical perturbation. Following
\citet{spiegel60} and \citet{chandrasekhar61}, the dynamics of a rotating Boussinesq system in Cartesian coordinates may
be ascertained from a parametric expansion of the Navier-Stokes equation under the assumption that quantities are scaled
relative to and expanded in the parameter $\alpha_T \Delta T$. This is equivalent to requiring that the maximum density
fluctuation be small in the sense that $\alpha_T\Delta T\ll 1$.  Under these assumptions, the equations of motion become

\vspace{-0.25truein}
\begin{center}
  \begin{align}
    \!\!\left(\dtime - \nu\nabla^2\right)\vvb &= -\vvb\bcdot\grad{\vvb}-\frac{1}{\langle\rho\rangle}\grad P' + g\alpha_TT'\zhat - 2\bomega\!\cross\!\vvb,\label{eqn:vv}\\
    \!\!\left(\dtime - \kappa\nabla^2\right) T' &= \beta \vcz - \vvb\bcdot\grad{T'},\label{eqn:Tprime}
  \end{align}
\end{center}

\noindent where $t$ is the time coordinate and $\partial_t$ is the partial derivative with respect to it, $\grad$ is the
spatial coordinate gradient. To first order in $\alpha_T \Delta T$, the continuity equation requires that the velocity
field $\vvb$ be solenoidal ($\dvg{\vvb}=0$). The vertical direction $\zhat$ is anti-aligned to the local direction of
the constant effective gravity $g$, e.g.  $\zhat=-\mathbf{g}/g$, so that fluid with $T'>0$ rises when $\Delta T<0$. As the
Cartesian f-plane domain assumed here is meant to correspond to a small, local region in the global spherical geometry,
the $x$ direction corresponds to the azimuthal direction and the $y$ direction to the latitudinal direction (See Figure
\ref{fig:coords}). Here $\langle\rho\rangle$ is the mean density, $P'$ is the pressure perturbation, $T'$ is the
temperature perturbation, $\vcz$ is the vertical component of the velocity field, $\bomega$ is the local rotational
vector that is taken to be in the y-z plane, and $\kappa$ and $\nu$ are the thermal and viscous diffusivities.
The vertically-aligned thermal gradient to first order in the expansion parameter $\alpha_T \Delta T$ is defined as
$\beta=d\bar{T}/dz+g/c_P$, where $c_P$ is the specific heat capacity at constant pressure.  The adiabatic lapse rate
$g/c_p$ arises from the first order expansion of the $P dV$ work in the thermal energy equation.

To construct a dispersion relationship for later use in the heat-flux maximization turbulence model, one may take the
$z$ component of both the curl and the double curl of Equation \ref{eqn:vv}. Dropping the nonlinear terms, these
operations yield the following set of linearized Boussinesq equations

\vspace{-0.25truein}
\begin{center}
  \begin{align}
    &\left(\dtime - \kappa\nabla^2\right) T' = \beta \vcz, \label{eqn:LTprime}\\
    &\left(\dtime - \nu\nabla^2\right)\zeta             = 2\bomega\cnabla{\vcz}, \label{eqn:zeta}\\
    &\left(\dtime - \nu\nabla^2\right)\nabla^2\vcz = g\alpha_T\nabla_{\perp}^2T' - 2\bomega\cnabla{\zeta},\label{eqn:d2vcz}
  \end{align}
\end{center}

\noindent with $\zeta$ being the vertical component of the vorticity and $\nabla_\perp$ being the gradient transverse to
the vertical direction as in Equations 79, 84, and 85 of \citet{chandrasekhar61}. As seen in many papers regarding
Boussinesq dynamics \citep[e.g.,][]{chandrasekhar61}, this set of equations can be reduced to a single third-order in
time and eighth-order in space equation for the vertical velocity as

\vspace{-0.25truein}
\begin{center}
  \begin{align}
    &\left(\dtime - \kappa\nabla^2\right)\left(\dtime - \nu\nabla^2\right)^2\nabla^2\vcz +
      g\alpha_T\beta\nabla_{\perp}^2\left(\dtime - \nu\nabla^2\right)\vcz \nonumber \\
           &\qquad+ 4\bomega\cnabla{\left[\bomega\cnabla{\left(\dtime - \kappa\nabla^2\right) \vcz}\right]} = 0. \label{eqn:fullvcz}
  \end{align}
\end{center}

For impenetrable and stress-free boundary conditions the solutions of Equation \ref{eqn:fullvcz} are periodic in the
horizontal, sinusoidal in the vertical, and exponential in time, e.g.
$\vcz = \rmv \sin{\left[k_z\left(z-z_c\right)\right]}\exp{(i\vkb_\perp\bcdot\vrb + s t)}$, where $\vkb_\perp$ is the
horizontal wavevector, $s$ is the growth rate, $\vrb$ is the local coordinate vector, and $\rmv$ is a to be determined
constant velocity amplitude. To satisfy the impenetrable, stress-free, and fixed temperature boundary conditions, it is required that
the vertical wavenumber be $k_z=n\pi/\ell_0$. This can also be considered to be equivalent to modeling only the portion
of the spectrum of convective eddies or motions whose Mach numbers are small and those that are much smaller than the
integral scale of the system. The introduction of this solution into Equation \ref{eqn:fullvcz} yields the following
dispersion relationship that relates $s$ to the wavevector $\vkb$ as

\vspace{-0.25truein}
\begin{center}
  \begin{align}
    &\left[\left(s +\kappa k^2\right)\left(s + \nu k^2\right)^2 k^2 + 
            g\alpha_T\beta k_{\perp}^2\left(s + \nu k^2\right)\right. \nonumber \\
    &\qquad\left.+ 4\left(\bilinear{\bomega}{\vkb}\right)^2\left(s + \kappa k^2\right)\right] \vcz = 0. \label{eqn:fvcz}
  \end{align}
\end{center}

\noindent This equation can be simplified and made non-dimensional by dividing through by the appropriate powers of $N_*$
and $k_z$, where $N_*^2=|g\alpha_T\beta|$ is the absolute value of the square of the Brunt-V\"{a}is\"{a}l\"{a} frequency
as in \citet{barker14}, which is otherwise negative in a convective region.  Utilizing the following auxiliary quantities

\vspace{-0.25truein}
\begin{center}
  \begin{align}
    \hat{s} &=\frac{s}{N_*}, \,\,\,
    z^3 = 1+ a^2 = \frac{k^2}{k_z^2}, \,\,\,
    a^2 = \frac{k_x^2}{k_z^2}+\frac{k_y^2}{k_z^2} = a_x^2 + a_y^2, \label{eqn:defszakv}\\
    K &= \frac{\kappa k_z^2}{N_*}, \quad
    V = \frac{\nu k_z^2}{N_*}, \nonumber
  \end{align}
\end{center}

\noindent the dispersion relationship becomes

\vspace{-0.25truein}
\begin{center}
  \begin{align}
    &\left(\hat{s} \!+\! K z^3\right)\!\! \left(\! z^3\!\left(\hat{s}\!+\! V z^3\right)^2 \!\!+\! O^2 \left(\cos{\theta}
      + a_y\sin{\theta}\right)^2\!\right) \nonumber\\
    &-\! \left(z^3\!-\! 1\right)\!\!\left(\hat{s}\!+\! V z^3\right)\!=\!0, \label{eqn:fullchar}
  \end{align}
\end{center}

\noindent where $4(\bomega\cdot\vkb)^2/N_*^2= k_z^2 O^2 \left(\cos{\theta} + a_y\sin{\theta}\right)^2$ with

\vspace{-0.25truein}
\begin{center}
  \begin{align}
    O^2 &= \frac{4\Omega_0^2}{N_*^2},
  \end{align}
\end{center}

\noindent where $\Omega_0$ is the bulk rotation rate of the system.

In assessing the behavior of the Boussinesq heat flux, both with and without diffusion, there are several limits of the
diffusive parameters that can be considered, each with its own physical justification.  In highly turbulent regimes, it
is often assumed that both diffusivities can be neglected, which is the case considered in \citetalias{stevenson79}, and
expanded upon below in \S\ref{sec:dfree}. Within stellar interiors, the molecular value of the thermal Prandtl number
($\mathrm{Pr}=\nu/\kappa$) is typically very small, being of the order of $10^{-5}$ or smaller, so the limit
$\nu\rightarrow 0$ may be considered as in \S\ref{sec:thermaldiff}.  The case where thermal diffusion can be ignored
($\kappa\rightarrow 0$) can be relevant to very high Prandtl number systems such as those found in geophysical contexts,
but it is not directly considered here. However, since the general case with both diffusivities can be treated here
(\S\ref{sec:viscdiff}), turbulent diffusivities may also be incorporated into the model, where typically it is assumed
that $\mathrm{Pr}$ is approximately unity. Moreover, to make contact with numerical simulations and laboratory
experiments, it is useful to retain diffusive effects.

\subsection{Maximum Heat Transport}\label{sec:mheat}

The secular impacts of rotation and magnetic fields on stellar and planetary evolution are of keen interest within the
astrophysical community \citep[e.g.,][]{maeder09,mathis13a}. Accordingly, extensions to MLT and its ilk have been
introduced \citep[e.g.,][]{stevenson79,canuto86,canuto91,zhou95,xiong97}. As expounded upon in \citet[][hereafter
S79]{stevenson79}, a surprisingly effective approach to including rotation in MLT is to hypothesize a convection model
where the convective length-scale, degree of superadiabaticity, and velocity are governed by the linear mode that
maximizes the convective heat flux. This heuristic method was made more rigorous when incorporated into the spectrally
nonlocal large-scale turbulence model of \citet{canuto86}, where it is used to set the scale of energy injection of a
Heisenberg-Kolmogorov spectrum of turbulent energy cascade \citep{kolmogorov41,heisenberg48}. Subsequently, it sets the
local values of the turbulent diffusivities. Though, these theories have not yet widely been employed in stellar
evolution computations, with one exception being \citet{ireland18}.  One can also incorporate approximate 2D dynamics
through models of the Reynolds and Maxwell-stresses, which has been significantly developed over the last several
decades \citep[e.g.,][]{durney79,hathaway84,ruediger89,garaud10}.

The \citetalias{stevenson79} model of rotating convection has its origins in the principle of maximum heat transport
proposed by \citet{malkus54}.  In that principle, an upper limit for a boundary condition dependent turbulent heat flux
is established that depends upon the smallest Rayleigh unstable convective eddy. The size of this eddy is determined
with a variational technique that is similar to that developed in \citet{chandrasekhar61} for the determination of the
Rayleigh number, which is the ratio of the buoyancy force to the viscous force multiplied by the ratio of the thermal to
viscous diffusion timescales. This technique then permits the independent computation of the rms values of the
fluctuating temperature and velocity amplitudes. Yet the heat-flux maximization principle is not fully rigorous as there
is no apriori theoretical justification as to why the turbulence would arrange itself so as to maximize the heat flux
\citep[e.g.,][]{howard63}. It is instead built upon three assumptions: first the mean-temperature gradient must be
everywhere negative, second there is a finite range of wavenumbers that are effective at transporting heat, and third
that the highest vertical wavenumber contributing to the heat transport is marginally stable with respect to a given
mean temperature profile. As \citet{howard63} has shown, however, the first assumption can be relaxed if the power
integrals of the Boussinesq equations are considered in place of the Boussinesq equations. It was also shown that there
will be a hierarchy of such constraint integrals that close the theory for each order of expansion.  So, it may be more
practical to appeal to a more mathematically complete means to compute the most probable equilibrium states, which is
the variational technique established in \citet{prigogine65}. Indeed, as \citet{spiegel62,townsend62} and
\citet{howard63} all point out, these optimal solutions are formed from solutions to the linear equations and hence they
are not true solutions to the full nonlinear Boussinesq equations. However, this flaw can be mollified if, as in
\citetalias{stevenson79}, it is conjectured that the convective modes are amplitude limited by instabilities in a
nonlinearly saturated state.

The Rayleigh-B\'{e}nard experiments carried out in \citet{townsend59} aim to directly test the heat-flux maximization
principle of \citet{malkus54}, where the temperature and its gradient are measured to compute the heat flux and the
moments of the temperature distribution. These values are then compared against the mean temperature gradient predicted
in \citet{malkus54}.  \citet{townsend59} finds that indeed the measured mean temperature gradient is well-described by
those expected from the heat-flux maximization principle, being inversely proportional to the vertical coordinate
outside of the boundary layer. This is in contrast to the self-similarity solution of \citet{priestley54}, which
predicts a scaling proportional to the inverse cube root of the vertical coordinate. \citet{howard63} reexamines those
experimental data with an emphasis on testing wether or not both the mean-fields and the fluctuating fields predicted in
\citet{malkus54} correspond to the measured flow structures, which are both found to match the theoretical values to
within a factor of order unity. Additionally, \citet{howard63} constructs a variational technique to assess the Rayleigh
number ($Ra = -g\alpha_T\Delta T \ell_0^3/\nu\kappa$) and the Nusselt number (which measures the ratio of the total heat
flux to the rate of thermal diffusion) of the heat-flux maximizing flows. It is also shown that, when the continuity
equation is imposed in addition to the power integrals, the experimentally measured Nusselt number is asymptotically
well-approximated by the variationally obtained value although its magnitude is overestimated by about a factor of
two. This approach can also be taken to compare Malkus's theory to more recent laboratory and numerical experiments of
Rayleigh-B\'{e}nard convection \citep[e.g.,][]{funfschilling05,ahlers06,anders17}. Furthermore, similar experiments of
Rayleigh-B\'{e}nard convection in rotating cylinders have measured the Nusselt number scaling with respect to the Rayleigh
number and with the Ekman number [$\mathrm{Ek}=\nu/(2\Omega\ell_0^2)$] in the rotating cases
\citep{zhong09,king09,king12a,king12b,king13,cheng15,aurnou15,weiss16}. Thus, a Malkus-like turbulence theory that
includes rotation, as in \citetalias{stevenson79} and as shown here, can in principle also be compared to those
experiments, which will be a focus of later work.

Laboratory experiments \citep{coates97,weiss16} and numerical simulations have lent some credence to this convection model
\citep{kapyla05,barker14,sondak15}. In particular, those simulations indicate that the low convective Rossby number scaling
regime established in \citetalias{stevenson79} appears to hold up well for three decades in convective Rossby number and
for about one decade in Nusselt number.  Moreover, a recent implementation of Stevenson's asymptotic scaling regimes in
convective Rossby number have been used to provide a local model of a rotationally-dependent mixing-length theory
\citep{ireland18}. There, it is shown that the entropy gradient of fully convective stars can be significantly modified
by rapid rotation in the bulk of their interiors, excluding the surface region where the convective Rossby number is
large.  Therefore, the adiabat of the star is modified, leading to structural and evolutionary changes. What remains to
be shown is how such a model of convection can impact the mixing for more modest rotation rates where the convective
Rossby numbers are closer to unity, the depth of convective penetration, as well as the amplitude of wave-driven and
shear-induced transport mechanisms.

\subsection{General Framework}\label{sec:genframe}

Following \citetalias{stevenson79}, the nonlinear saturation conjecture used in conjunction with the maximum heat
transport principle requires that the primary contribution to each scale of the temperature and velocity fluctuation is
from the convective term at the corresponding scale. This implies that its growth rate is
$s = \left|\vvb\right| \left|\vkb\right|$, which also provides a way to estimate the heat flux
$F = \rho c_P \Re{[\vcz T']}$ with linear dynamics. Moreover, in the theory of Malkus and Howard, the mode that carries
the largest fraction of the heat flux is the fundamental vertical mode, e.g. $n=1$ so that $k_z = \pi/\ell_0$, but whose
horizontal structure is undetermined.  It is this mode that Stevenson's turbulence model is built around.  Nominally,
however, the maximization should be carried out scale-wise for each of the modes that contribute to the heat flux. This
task will be left for later work that examines non-Boussinesq turbulence using Malkus's theory as a point of
comparison. Thus, Equation \ref{eqn:vv}, with the assumptions used above to formulate the dispersion relationship, yields

\vspace{-0.25truein}
\begin{center}
  \begin{align}
    &s \vvb = -i\vkb P'/\langle\rho\rangle +g\alpha_TT'\zhat -2\bomega\cross\vvb - \nu k^2\vvb, \label{eqn:fvv}\\
    &i\bilinear{\vkb}{\vvb} = 0.\label{eqn:vsol}
  \end{align}
\end{center}

\noindent Taking the dot product of $\vvb$ and Equation \ref{eqn:fvv}, using Equation \ref{eqn:vsol} to eliminate
the pressure, and taking a volume average it can be seen that

\vspace{-0.25truein}
\begin{center}
  \begin{align}
    \left(s+\nu k^2\right) \langle\rmv^2\rangle = g\alpha_T\langle T'\vcz\rangle. \label{eqn:v_theta_link}
  \end{align}
\end{center}

\noindent Therefore,

\vspace{-0.25truein}
\begin{center}
  \begin{align}
    F = \frac{\rho c_P\Re{[\left(s+\nu k^2\right) \langle\rmv^2\rangle]}}{g\alpha_T}. 
  \end{align}
\end{center}

\noindent Employing the hypothesis that the velocity in the quasi-stationary turbulent state scales as
$\langle\rmv^2\rangle = s s^{*}/k^2$, one has the heat flux

\vspace{-0.25truein}
\begin{center}
  \begin{align}
    F &= \frac{\langle\rho\rangle c_P}{g\alpha_T} \frac{\Re{[\left(s+\nu k^2\right) ]}s s^{*}}{k^2}\\
      &= \frac{\langle\rho\rangle c_P N_*^3}{g\alpha_T k_z^2}\left[\frac{\Re{[\hat{s}]}}{z^3}+V\right]\left|\hat{s}\right|^2, \label{eqn:maxhf}
  \end{align}
\end{center}

\noindent where the definitions given in Equations \ref{eqn:defszakv} have been used to simplify the latter expression.

It is useful to eliminate some of the possible solutions to the maximization of this heat flux. First, consider the
nondiffusive case, where

\vspace{-0.25truein}
\begin{center}
  \begin{align}
    \hat{s}^2 = \frac{a^2 - O^2\left(\cos{\theta}+\sin{\theta}a_y\right)^2}{1+a^2},
  \end{align}
\end{center}

\noindent from which it is clear that any value of $a_y$ will reduce the growth rate.  This also places a lower limit on
the value of $a$ for which the solutions are real with a growing branch, with $a_x>O\cos{\theta}$, which is shown to
be satisfied by the heat-flux maximization in \citet{stevenson79} and in the next subsection.  Likewise, in the
nonrotating but diffusive case, one has that

\vspace{-0.25truein}
\begin{center}
  \begin{align}
   \left[\hat{s}+V z^3\right]\left[z^3\left(\hat{s}+K z^3\right)\left(\hat{s}+V z^3\right)-a^2\right]=0.
  \end{align}
\end{center}

\noindent This dispersion relationship has a growing real solution for sufficiently small $V$ and $K$ as

\vspace{-0.25truein}
\begin{center}
  \begin{align}
    \hat{s} &= z^{-3}\sqrt{\left(V+K\right)^2z^{12}+4 z^3\left(a^2-K V z^9\right)}\nonumber\\
    &-\frac{1}{2}\left(V+K\right)z^3.
  \end{align}
\end{center}

\noindent When linearized with respect to $V$ and $K$, this yields

\vspace{-0.25truein}
\begin{center}
  \begin{align}
    \hat{s} &= a z^{-\frac{3}{2}}-\frac{1}{2}\left(V+K\right)z^3,
  \end{align}
\end{center}

\noindent which is maximized when $a_y=0$. So, in all the instances of parameter regimes considered below, the
maximization will be carried out over $z^3=1+a_x^2$ with $a_y=0$.  Note that this is equivalent to considering only the
two-dimensional rolls aligned with the rotation axis, which in the rotating case is due largely to the
Taylor-Proudman constraint.  These modes also have a larger growth rate than three-dimensional modes for
non-rotating Rayleigh-B\'{e}nard convection, but nevertheless the 2D modes can approximate many of the behaviors of the
3D modes \citep[e.g.,][]{vanderpoel13}.

It is also useful to recast some of the parameters so that the system can be more readily compared to the
\citetalias{stevenson79} results. The characteristic velocity $\rmv_0$ is derived from the growth rate and maximizing
wavevector in the nonrotating and nondiffusive case, which are $s_0^2=3/5|g_0\alpha_T\beta_0|=3/5N^2_{*,0}$, with
$\beta_0$ being the thermal gradient and $g_0$ being the effective gravity in the nonrotating case, and
$k_0^2=5/2 k_z^2$ as in \citep{stevenson79} Equation 36, leading to

\vspace{-0.25truein}
\begin{center}
  \begin{align}
    \rmv_0 = \frac{s_0}{k_0} = \frac{\sqrt{6}}{5} \frac{N_{*,0}}{k_z} =  \frac{\sqrt{6}}{5\pi} N_{*,0}\ell_0.\label{eqn:v0}
  \end{align}
\end{center}

\noindent This permits the definition of the convective Rossby number $\Roc$, which for this Boussinesq system is

\vspace{-0.25truein}
\begin{center}
  \begin{align}
    \Roc=\frac{\rmv_0}{2\Omega_0 \ell_0}=\frac{\sqrt{6} N_{*,0}}{10\pi\Omega_0}, \label{eqn:rossby}
  \end{align}
\end{center}

\noindent which implies that

\vspace{-0.25truein}
\begin{center}
  \begin{align}
    O &=\frac{2\Omega_0}{N_*} = \frac{\rmv_0}{N_* \Roc\ell_0} = \frac{\sqrt{6}N_{*,0}}{5\pi N_*\Roc}. \label{eqn:o0def}
  \end{align}
\end{center}

Next consider the variation of the superadiabaticity, which for this system is given by $\epsilon = H_P \beta/T$,
meaning that $N_*^2 = |g \alpha_T T \epsilon/H_P|$, where $H_P$ is the pressure scale height.  The comparison made
throughout this paper is between cases with additional included physical effects and the base case that is nondiffusive
and nonrotating.  The potential temperature gradient in this latter case is ascertained from the Malkus-Howard
turbulence model, which yields a value of $N_{*,0}$.  So far, all quantities have been normalized with respect to
$N_*$. Instead, it is useful to compare them relative to $N_{*,0}$ and to introduce the ratio of superadiabaticities
as an additional unknown as

\vspace{-0.25truein}
\begin{center}
  \begin{align}
	q &= N_{*,0}/N_*.\label{eqn:qdef}
  \end{align}
\end{center}

\noindent Therefore, all parametric quantities have the following equivalencies

\vspace{-0.25truein}
\begin{center}
  \begin{align}
    O &= q \frac{\sqrt{6}}{5\pi\Roc}= q O_0,\nonumber\\
    K &= q \frac{\kappa k_z^2}{N_{*,0}} = q K_0,\label{eqn:equivalencies}\\
    V &= q \frac{\nu k_z^2}{N_{*,0}} = q V_0.\nonumber
  \end{align}
\end{center}

So, the dispersion relationship (Equation \ref{eqn:fullchar}) and the heat flux (Equation \ref{eqn:maxhf}) may be
written as

\vspace{-0.25truein}
\begin{center}
  \begin{align}
    &\left(\hat{s} \!+\! K_0 q z^3\right)\!\! \left(\! z^3\!\left(\hat{s}\!+\! V_0 q z^3\right)^2 \!\!+\! O_0^2q^2\cos^2{\theta}\!\right) \!\nonumber\\
    &\qquad-\! \left(z^3\!-\! 1\right)\!\!\left(\hat{s}\!+\! V_0 q z^3\right)\!=\!0, \label{eqn:fullcharq}\\
    &F = \frac{F_0}{q^3} \left[\frac{\hat{s}^3}{z^3}+V_0 q\hat{s}^2\right],\label{eqn:heatfluxq}
  \end{align}
\end{center}

\noindent where $F_0 = \langle\rho\rangle c_P N_{*,0}^3/\left(g\alpha_T k_z^2\right)$. When maximized for a positive real value of
$\hat{s}$, Equation \ref{eqn:heatfluxq} implies that

\vspace{-0.25truein}
\begin{center}
  \begin{align}
      \frac{d\hat{s}}{dz} = \frac{3\hat{s}^2}{z\left(3\hat{s}+2V_0 q z^3\right)}. \label{eqn:maxsigma}
  \end{align}
\end{center}

Depending upon the method of finding $\hat{s}$, it can be quite complicated to compute directly as a function of $z$.
Instead, it is useful to compute the implicit derivative of the dispersion relationship (Equation \ref{eqn:fullcharq})
to obtain $d\hat{s}/dz$, which yields

\vspace{-0.25truein}
\begin{center}
  \begin{align}
    &\frac{d\hat{s}}{dz}=-3z^2\left[V_0q\left(1-2z^3\right)+K_0 q^3\left(O_0^2\cos^2{\theta}+4V_0^2z^9\right) \right. \nonumber\\
    &\left.+ \hat{s}\left(\hat{s}^2+2q\left(K_0+2V_0\right) z^3 \hat{s} + 3 V_0 q^2\left(2K_0+V_0\right)z^6-1\right)\right]\nonumber\\
    &\left[z^3\!\left(3\hat{s}^2\!+\! 2q\left(K_0\!+\! 2V_0\right)z^3\hat{s}\!+\! V_0q^2\left(2K_0\!+\! V_0\right)z^6\!-\! 1\right)\right.\nonumber\\
    &\left.+ q^2 O_0^2\cos^2{\theta}\!+1\right]^{-1}\!\!\!\!. \label{eqn:dsigma}
  \end{align}
\end{center}

Finally, to assess the scaling of the superadiabaticity, the velocity, and the horizontal wavevector, a further
assumption must be made in which the maximum heat flux is invariant to any parameters, namely that
$\max{\left[F\right]}=\max{\left[F\right]}_0$ so the heat flux is equal to the maximum value $\max{\left[F\right]}_0$
obtained in the Malkus-Howard turbulence model for the nonrotating case.  This is based on the assumption that the total
heat flux should not change with rotation, which is not true in general as the centripetal acceleration of the star can
lead to a variation in the central temperature and density of the star and thus to a variation of the total luminosity
of the star at a fixed mass \citep[e.g.,][]{rieutord16}.  Furthermore, the flux will be latitude dependent due to the
rotational variation of the local gravity \citep[e.g.,][]{maeder99,wang16}. Indeed, an analysis of 3D nonlinear
global-scale convection simulations of rapidly rotating oblate stars has been carried out in \citet{wang16}.  These
simulations indicate that the Von Zeipel theorem \citep{vonzeipel24} holds well even for the emergent flux at the outer
boundary, with the exception of a region near the equator where there is a flux enhancement. Thus, while acknowledging
that the flux derived from the Von Zeipel theorem is not rigorously representative of the heat flux in a convection
zone, it provides a very good approximation. Following \citet{maeder99} and the von Zeipel theorem, one may show that
the heat flux $F_V$ on an isobar with $P=\langle P\rangle$ varies as

\vspace{-0.25truein}
\begin{center}
  \begin{align}
    F_V = \frac{L\!\left(\langle P\rangle\right)}{4\pi G M_*\!\!\left(\langle P\rangle\right)} g,
  \end{align}
\end{center}

\noindent where $L\!\left(\langle P\rangle\right)$ is the total luminosity on the isobar and
$M_*\!\left(\langle P\rangle\right)$ is the total mass integrated to the isobar. Approximating the effective gravity and the
shift in the mass as in \citet{maeder99} (Equation 2.3), this becomes

\vspace{-0.25truein}
\begin{center}
  \begin{align}
    F_V \approx \frac{L\!\left(\langle P\rangle\right) \left(g_0-\Omega_0^2r\sin{\theta}\right)}{4\pi G M\left[1-\Omega_0^2/\left(2\pi G \langle\rho\rangle\right)\right]},
  \end{align}
\end{center}

\noindent where $r$ and $\theta$ correspond to the origin of the local coordinate system as in Figure
\ref{fig:coords}. Taking the ratio of the rotating and non-rotating fluxes, one has that

\vspace{-0.25truein}
\begin{center}
  \begin{align}
    \frac{F_V}{F_{V,0}} \approx \frac{L\!\left(\langle P\rangle\right)}{L\left(r\right)}\frac{1-\Omega_0^2r\sin{\theta}/g_0}{1-\Omega_0^2/\left(2\pi G\langle\rho\rangle\right)},
  \end{align}
\end{center}

\noindent which in terms of the convective Rossby number and the buoyancy frequency is

\vspace{-0.25truein}
\begin{center}
  \begin{align}
    \frac{F_V}{F_{V,0}} \approx \frac{L\!\left(\langle P\rangle\right)}{L\left(r\right)}
    \frac{1-3N_{*,0}^2r\sin{\theta}/\left(50\pi^2 \mathrm{Ro_c}^2 g_0\right)}{1-3 N_{*,0}^2/\left(100\pi^3 \mathrm{Ro_c}^2 G\langle\rho\rangle\right)}.
  \end{align}
\end{center}

\noindent Noting the definition of the buoyancy frequency amplitude and the local density, one can see that

\vspace{-0.25truein}
\begin{center}
  \begin{align}
    &\frac{F_V}{F_{V,0}} \approx \frac{L\!\left(\langle P\rangle\right)}{L\left(r\right)}
    \frac{1-3|\alpha_T\Delta T| \sin{\theta}/\left(50\pi^2 \mathrm{Ro_c}^2 \right)}{1-|\alpha_T\Delta T|g_0
    r^2/\left(25\pi^2\mathrm{Ro_c}^2 G M\right)}\\
    &\approx \frac{L\!\left(\langle P\rangle\right)}{L\left(r\right)} \left[1+\frac{|\alpha_T \Delta
      T|}{25\pi^2\mathrm{Ro_c}^2}\left(\frac{g_0r^2}{GM}-\frac{3}{2}\sin{\theta}\right)\right] +\cdots.
  \end{align}
\end{center}

Therefore, since the linearized change in flux is proportional to the Boussinesq expansion parameter
$|\alpha_T \Delta T|$ and the inverse square of the convective Rossby number, this requires
$\mathrm{Ro_c} > 5^{-1} \pi^{-1} |\alpha_T \Delta T|^{1/2}$ for the approximations of the model to hold.  Thus, for now,
these effects will be ignored as they are sensitive to the global-scale dynamics, microphysics of the nuclear energy
generation rate, and the equation of state. Therefore, building this convection model consists of three steps: defining
a dispersion relationship that links $\hat{s}$ to $q$ and $z$, maximizing the heat flux with respect to $z$, and
assuming an invariant maximum heat flux that then closes this three variable system.

In the case of planetary and stellar interiors, the viscous damping timescale is generally longer than the convective
overturning timescale (e.g., $V_0\ll N_{*,0}$).  Thus, the maximized heat flux invariance is much simpler to treat.  In
particular, the heat flux invariance condition under this assumption is then

\vspace{-0.25truein}
\begin{center}
  \begin{align}
    \frac{\max{\left[F\right]}}{\max{\left[F\right]}_0}
    &=\frac{25}{6}\sqrt{\frac{5}{3}}\left[\frac{\hat{s}^3}{q^3z^3}+\frac{V_0\hat{s}^2}{q^2}\right]_{\mathrm{max}}\nonumber\\
    &\approx\left.\frac{25}{6}\sqrt{\frac{5}{3}}\frac{\hat{s}^3}{q^3z^3}\right|_{\mathrm{max}} =1,
  \end{align}
\end{center}

\noindent implying that

\vspace{-0.25truein}
\begin{center}
  \begin{align}    
    &\hat{s}=\tilde{s} q z + \mathcal{O}(V_0),\label{eqn:maxndhf}
  \end{align}
\end{center}

\noindent where $\tilde{s}=2^{1/3} 3^{1/2} 5^{-5/6}$ and $\max{\left[F\right]}_0=6/25\sqrt{3/5} F_0$ follows from the
definition of the flux and the maximizing wavevector used to define $\rmv_0$ above in Equation \ref{eqn:v0}.

The assumption of this convection model is that the magnitude of the velocity is defined as the ratio of the maximizing
growth rate and wavevector. With the above approximation, the velocity amplitude can be defined relative to the
nondiffusive and nonrotating case scales without a loss of generality as

\vspace{-0.25truein}
\begin{center}
  \begin{align}
    \frac{\rmv}{\rmv_0} &= \frac{k_0}{s_0}\frac{s}{k} = \frac{5}{\sqrt{6}} \frac{N_*}{N_{*,0}}\frac{\hat{s}}{z^{3/2}} =
                          \frac{5}{\sqrt{6}} \frac{\hat{s}}{q z^{3/2}} = \left(\frac{5}{2}\right)^{\frac{1}{6}} z^{-\frac{1}{2}}.\label{eqn:vsteve}
  \end{align}
\end{center}

\noindent So only the maximizing wavevector needs to be found in order to ascertain the relative velocity amplitude. For
reference, the symbols that will be frequently used from this section are listed in Table \ref{tab1}.

\begin{table}
  \begin{center}
    \begin{tabular}{Ll}
      \hline
      \hline
      $a=k_x/k_z$ & Maximizing horizontal wavevector\\
      $k_z=\pi/\ell_0$ & Maximizing vertical wavevector \\
      $K_0=\kappa k_z^2/N_{*,0}$ & Normalized thermal diffusivity\\
      $O_0=\sqrt{6}/\left(5\pi\Roc\right)$ & Normalized Coriolis coefficient\\
      $\Roc=\sqrt{6}N_{*,0}/\left(10\pi\Omega_0\right)$ & Convective Rossby number\\
      $\hat{s}=s/N_*$ & Normalized growth rate\\
      $\rmv_0=\sqrt{6}N_{*,0}\ell_0/\left(5\pi\right)$ & Velocity of the nonrotating case\\
      $V_0= \nu k_z^2/N_{*,0}$ & Normalized viscosity\\
      $q=N_{*,0}/N_*$ & Ratio of buoyancy timescales\\
      $z^3 = k^2/k_z^2$ & Normalized wavevector\\
      \hline
    \end{tabular}
    \caption{Frequently used symbols in the convection model.}\label{tab1}
  \end{center}
\end{table}

\subsection{Diffusion-free Models}\label{sec:dfree}

As considered in \citetalias{stevenson79}, the simplest case to consider is one in which all diffusive mechanisms are
neglected. Given Equation \ref{eqn:maxsigma}, the maximization condition requires

\vspace{-0.25truein}
\begin{center}
  \begin{align}
   \frac{d\hat{s}}{dz} &= \frac{\hat{s}}{z},\label{eqn:maxwov}
  \end{align}
\end{center}

\noindent where the derivative of the growth rate can be determined from the nondiffusive limit of Equation \ref{eqn:dsigma} as 

\vspace{-0.25truein}
\begin{center}
  \begin{align}
   \frac{d\hat{s}}{dz} &= -\frac{3z^2\hat{s}\left(\hat{s}^2-1\right)}{1+O_0^2 q^2\cos^2{\theta}+z^3\left(3\hat{s}^2-1\right)};
  \end{align}
\end{center}

\noindent when equated this yields the following equation for the growth rate

\vspace{-0.25truein}
\begin{center}
  \begin{align}
  \hat{s}^2 = \frac{4z^3-O_0^2q^2\cos^2{\theta}-1}{6z^3}.
  \end{align}
\end{center}

\noindent However, the dispersion relationship (Equation \ref{eqn:fullcharq}) for the nondiffusive case requires

\vspace{-0.25truein}
\begin{center}
  \begin{align}
    \hat{s}^2 = \frac{z^3-O_0^2q^2\cos^2{\theta}-1}{z^3}.\label{eqn:ndcharq}
  \end{align}
\end{center}
    
\noindent Using the heat flux invariance condition (Equation \ref{eqn:maxndhf}) so that
$\hat{s}=\tilde{s} q z$, the two above equations can be manipulated to isolate a single equation for the
maximizing value of $z$ and a dependent equation for $q$ as

\vspace{-0.25truein}
\begin{center}
  \begin{align}
    &2z^5-5z^2-\frac{3}{\tilde{s}^2} O_0^2\cos^2{\theta} = 0,\\
    &q^2 - \frac{z^3-1}{\tilde{s}^2z^5+O_0^2\cos^2{\theta}}=0.
  \end{align}
\end{center}

\noindent Together with Equations \ref{eqn:equivalencies}, these two equations yield the following scalings with
convective Rossby number and maximizing wavevector as

\vspace{-0.25truein}
\begin{center}
  \begin{align}
    \frac{\epsilon}{\epsilon_0} &= \frac{25\pi^2\Ro_{\mathrm{c}}^2\tilde{s}^2z^5+6\cos^2{\theta}}{25\pi^2\Ro_{\mathrm{c}}^2\left(z^3-1\right)},\label{eqn:xsteve}\\
    2&z^5-5z^2-\frac{18\cos^2{\theta}}{25\pi^2\Ro_{\mathrm{c}}^2\tilde{s}^2}=0.\label{eqn:zsteve}
  \end{align}
\end{center}
    
\begin{center}
  \begin{figure}[t!]
    \hspace*{-0.25cm}\includegraphics[width=0.5\textwidth]{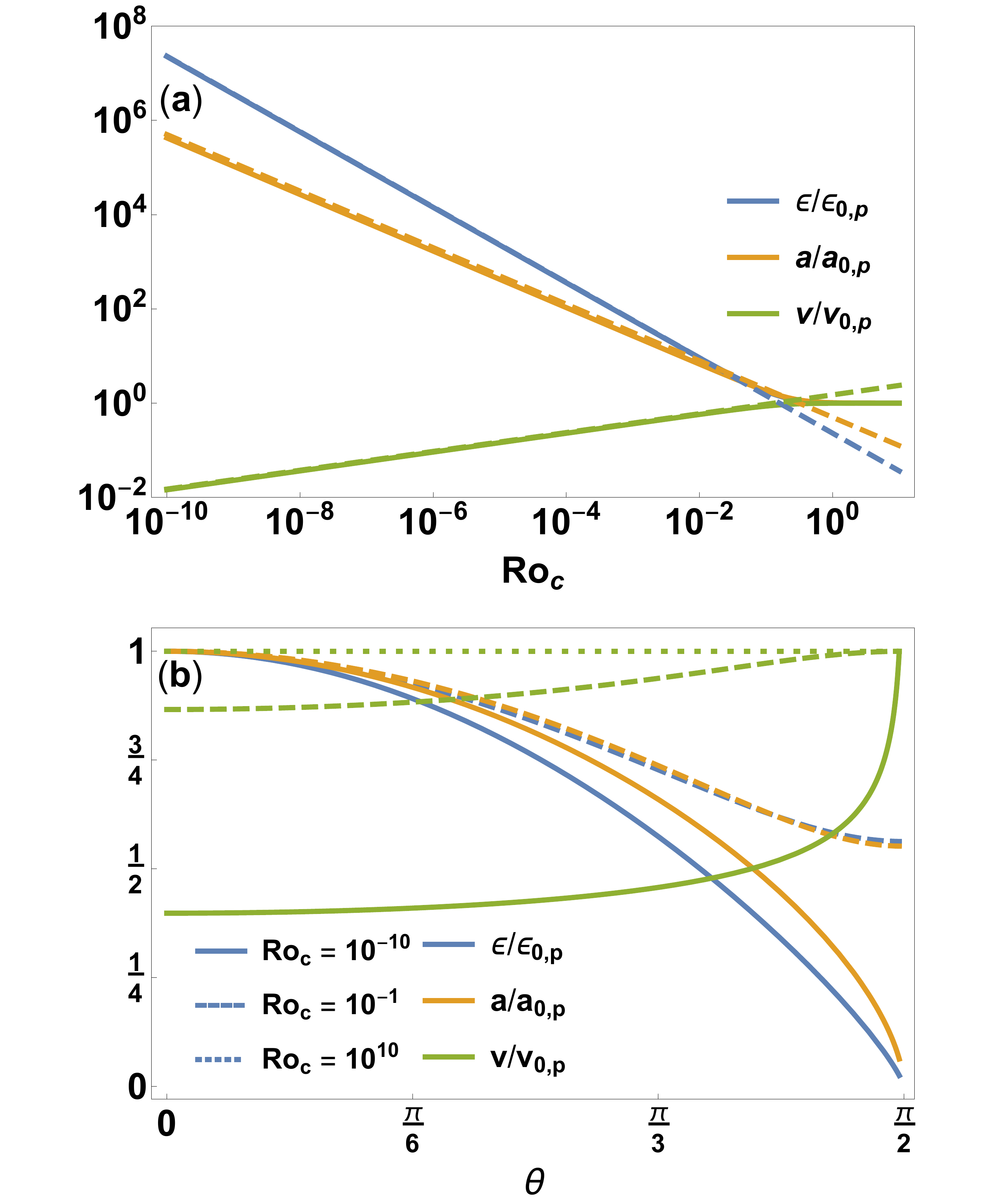} 
    \caption{Convective Rossby number and latitudinal dependence of the diffusion-free convection model.  (a) The
      superadiabaticity (blue), horizontal wavevector (orange), and velocity (green) dependence upon the convective
      Rossby number at the pole ($\theta=0$). The asymptotic scaling given in \citetalias{stevenson79} are shown as
      dashed lines. (b) Latitudinal dependence of the superadiabaticity (blue), horizontal wavevector (orange), and
      velocity (green) for the diffusion-free convection model at three extremal convective Rossby numbers $10^{-10}$
      (solid), $10^{-1}$ (dashed), and $10^{10}$ (dotted), with each case normalized by their value at the pole.  There
      is effectively no change with latitude at very large convective Rossby number, leading to overlapping dotted
      curves.}\label{fig:steve_scaling}
  \end{figure}
\end{center}

With the quintic form of Equation \ref{eqn:zsteve}, its solutions are not representable as radicals or other functions.
So, it is left in its current form to be solved numerically for a given colatitude and convective Rossby
number. However, at low convective Rossby number, these equations can be asymptotically approximated: for $\Ro_c\ll1$ it
is clear that $z \propto \Ro_{\mathrm{c}}^{-2/5}$ since $z^2\ll z^5$. It then follows that
$\epsilon/\epsilon_0 \propto \Ro_{\mathrm{c}}^{-4/5}$ and $\rmv/\rmv_0 \propto \Ro_{\mathrm{c}}^{1/5}$, all of which
match the scalings given in \citetalias{stevenson79}, the direct numerical simulations of \citet{kapyla05} and
\citet{barker14}, and can be seen in Figure \ref{fig:steve_scaling}(a).  In contrast, at very large convective Rossby
number, these quantities converge to unity as expected.

At a first glance, one can see from Equations \ref{eqn:xsteve} and \ref{eqn:zsteve} that at the equator there is no
rotational scaling of any of these quantities.  However, there is a rapid transition from this behavior just a few
degrees away from the equator.  At a fixed convective Rossby number, the superadiabaticity has a latitudinal dependence
of $\cos^2{\theta}$, meaning that its value at a larger colatitude is reduced with respect to its maximum value at the
pole. Therefore, in order to maintain a constant heat flux, the velocity must increase toward the equator.  In contrast,
the horizontal wavevector is also a monotonically decreasing function of colatitude, with a maximum at the pole and
minimum at the equator.  Hence, the heat carrying motions are of a larger scale at larger colatitudes than at the pole
as expected in theoretical considerations \citep[e.g.,][]{busse02,dormy04} and as seen in numerical simulations of
spherical rotating convection in both fully convective domains and in spherical shells
\citep[e.g.,][]{glatzmaier85,miesch05,brun05,browning08, augustson16}. These behaviors are illustrated in Figure
\ref{fig:steve_scaling}(b).

\subsection{Adding Thermal Diffusion}\label{sec:thermaldiff}

The procedure developed above can be applied again in the case with thermal diffusion as the heat flux maximization
condition remains unchanged from Equation \ref{eqn:maxwov}. Taking the limit that $V\rightarrow 0$ in Equation
\ref{eqn:dsigma}, one finds that 

\vspace{-0.25truein}
\begin{center}
  \begin{align}
    \frac{d\hat{s}}{dz} &\!=\!-\frac{3z^2\left(q^3 K_0 O_0^2\cos^2{\theta} +\hat{s}\left(\hat{s}^2+2 K_0 q z^3\hat{s}-1\right)\right)}{1+q^2 O_0^2\cos^2{\theta}+z^3\left(3\hat{s}^2+2K_0 q z^3\hat{s}-1\right)},
  \end{align}
\end{center}

\noindent which implies the following heat flux maximization condition

\vspace{-0.25truein}
\begin{center}
  \begin{align}
    &\hat{s}\left(1+q^2O_0^2\cos^2{\theta}+2z^3\left(3\hat{s}^2+4q K_0z^3\hat{s}-2\right)\right) \nonumber\\
    &\qquad+3q^3 K_0 O_0^2 z^3\cos^2{\theta} =0. \label{eqn:Kcond}
  \end{align}
\end{center}

\noindent The dispersion relationship (Equation \ref{eqn:fullcharq}) with $V_0\rightarrow 0$ is

\vspace{-0.25truein}
\begin{center}
  \begin{align}
    &\left(\hat{s} \!+\! K_0 q z^3\right)\!\! \left(\! \hat{s}^2 z^3\!+\! O_0^2q^2\cos^2{\theta}\!\right) \!-\! \hat{s}\left(z^3\!-\! 1\right)\!=\!0.\label{eqn:Kchar}
  \end{align}
\end{center}

\noindent Again utilizing the maximized heat flux constraint $\hat{s}=\tilde{s}qz$ and the definitions
in Equations \ref{eqn:equivalencies}, the two previous equations can be shown to be equivalent to

\vspace{-0.25truein}
\begin{center}
  \begin{align}
    &q^2 - \frac{\tilde{s}\left(z^3-1\right)}{\left(\tilde{s} + K_0 z^2\right)\left( \tilde{s}^2z^5 +O_0^2\cos^2{\theta}\right)} \!=\! 0,\\
    &\tilde{s}z\left(2z^5\!-\! 5z^2\!-\! \frac{3}{\tilde{s}^2} O_0^2\cos^2{\theta}\right)\nonumber\\
    &\qquad+\! K_0\left[4z^8\!-\! 7z^5\!-\!\frac{O_0^2}{\tilde{s}^2}\cos^2{\theta}\left(z^3\!+\! 2\right)\right]\!=\!0.
  \end{align}
\end{center}

\noindent So the connection to the nondiffusive case is clear and can be recovered directly when taking the limit as
$K_0\rightarrow 0$.  Employing the relationship between $O$ and the convective Rossby number given in Equation
\ref{eqn:o0def}, the superadiabaticity and wavevector can be defined as

\vspace{-0.25truein}
\begin{center}
  \begin{align}
    &\frac{\epsilon}{\epsilon_0} = \frac{\left(\tilde{s} + K_0 z^2\right)\left( 25 \pi^2 \Ro_{\mathrm{c}}^2 \tilde{s}^2z^5 +6 \cos^2{\theta}\!\right)}{25 \pi^2 \Ro_{\mathrm{c}}^2 \tilde{s}\left(z^3-1\right)},\label{eqn:xthermal}\\
    &\tilde{s} z\left[2z^5- 5z^2- \frac{18\cos^2{\theta}}{25\pi^2\Ro_{\mathrm{c}}^2\tilde{s}^2}\right]\nonumber\\
    &\qquad+K_0\left[4z^8-7z^5- \frac{6\cos^2{\theta}\left(z^3+ 2\right)}{25\pi^2\Ro_{\mathrm{c}}^2\tilde{s}^2}\right]= 0.\label{eqn:zthermal}
  \end{align}
\end{center}

\begin{figure}[t!]
  \begin{center}
    \hspace*{-0.5cm}\includegraphics[width=0.5\textwidth]{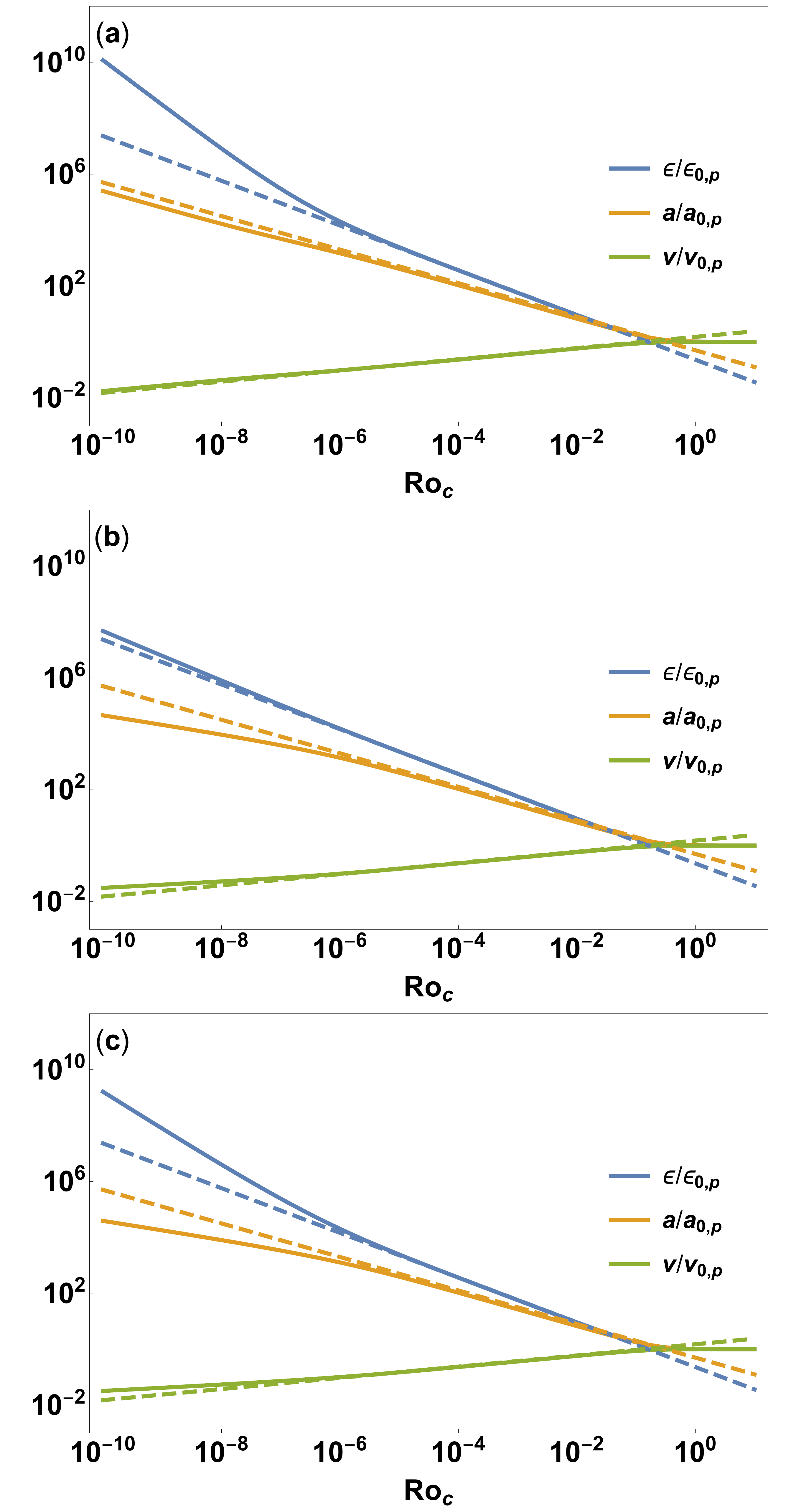} 
    \caption{Convective Rossby number dependence of the diffusive convection model at the pole ($\theta=0$), showing the scaling of
      the superadiabaticity (blue), the horizontal wavevector (orange), and the velocity (green). The asymptotic scaling
      given in \citetalias{stevenson79} are shown as dashed lines. (a) The thermally diffusive case ($K_0=10^{-5}$, $V_0=0$),
      (b) the viscous case ($K_0=0$, $V_0=10^{-5}$), and (c) the fully diffusive case ($K_0=10^{-5}$,
      $V_0=10^{-5}$).}\label{fig:diffusive_scaling}
  \end{center}
\end{figure}

Immediately, one can see that the superadiabaticity increases as the thermal diffusion is increased because the thermal
gradient must increase to drive convection.  The nondiffusive component is unchanged and so it also grows with lower
convective Rossby numbers.  The form of Equation \ref{eqn:zthermal} indicates that the wavevector must increase with decreasing
convective Rossby number, whereas its behavior with varying the thermal diffusion rate ($K_0$) is not obvious and is illustrated in
Figure \ref{fig:zscaling}(a) for the fully diffusive case.  There it is seen that thermal diffusion induces a shift in
the value of the wavevector ratio that is a factor of $(7/10)^{1/3}$, or approximately 13\%, lower for convective Rossby numbers
below the value of $K_0$.  For $K_0>1$, the asymptotic value of $z$ at large convective Rossby number is reduced by the same
factor, from $(5/2)^{1/3}$ to $(7/4)^{1/3}$, indicating that the horizontal scale of the flows becomes larger. The
velocity scales inversely with the horizontal wavevector; so, as before, the velocity must decrease with increasing
rotation rate.  These behaviors are shown for a typical value of the thermal diffusion in Figure
\ref{fig:diffusive_scaling}(a). The velocity decreases with lower convective Rossby number given that the wavevector increases,
with the same nontrivial behavior with the thermal diffusivity.

\subsection{Including Viscosity}\label{sec:viscdiff}

When including viscosity within this heuristic framework, the maximum heat flux should formally be derived from the
algebraic system formed by the dispersion relationship for $\hat{s}$ and from

\vspace{-0.25truein}
\begin{center}
  \begin{align}
    F/F_0 &= \tilde{s}^3\left[\frac{\hat{s}}{q^3z^3}+V_0 q\right]\hat{s}^2=1. 
  \end{align}
\end{center}

\noindent In particular, the above equation, Equation \ref{eqn:fullchar}, and Equation \ref{eqn:maxsigma} give

\vspace{-0.25truein}
\begin{center}
  \begin{align}
    &\hat{s}^3+\left(V_0 q\hat{s}^2-\tilde{s}^3\right) q^3 z^3=0,\\
    &\left(\hat{s}+K_0 q z^3\right)\left[z^3\left(\hat{s}+V_0 q z^3\right)^2+\frac{6\cos^2{\theta}}{25\pi^2\Ro_{\mathrm{c}}^2}\right]\nonumber\\
    &\qquad+\left(1-z^3\right)\left(\hat{s}+V_0 q z^3\right)=0, \\
    &\frac{6q^2\cos^2{\theta}}{25\pi^2\Ro_{\mathrm{c}}^2}\left[\hat{s}^2+K_0 q z^3\left(3\hat{s}+2 q V_0 z^3\right)\right]\nonumber\\
    &\qquad+\left(\hat{s}+V_0 q z^3\right)\left[6\hat{s}^3z^3 + 4 q \hat{s}^2 z^6\left(3V_0 + 2K_0\right)\right.\nonumber\\
    &\qquad +\hat{s}\left(1-4z^3+2V_0 q^2 z^9\left(8K_0+3V_0\right)\right)\nonumber\\
    &\qquad\left.+2V_0 q z^3\left(1-2z^3+4K_0V_0 q^2 z^9\right)\right]=0.
  \end{align}
\end{center}

While this set of equations can be solved numerically, they can be simplified in the case of planetary and stellar
interiors to a relatively simple scaling behavior with viscosity.  In particular, the viscous component of the heat flux
may be neglected if it is again assumed that $V_0\ll N_{*,0}$), so that as above
$\hat{s} \approx \tilde{s} q z + \mathcal{O}(V_0)$. Subsequently, following the method shown for the case
of thermal diffusion, one may find the implicit wavevector derivative of the growth rate $\hat{s}$ (Equation
\ref{eqn:dsigma}), which implies that the constraining dispersion relationship and the equation resulting from the heat
flux maximization are

\vspace{-0.25truein}
\begin{center}
  \begin{align}
    &\left(\tilde{s}+V_0 z^2\right)\left[q^2z^5\left(\tilde{s}+V_0z^2\right)\left(\tilde{s}+K_0z^2\right)-z^3+1\right]\nonumber\\
    &+\frac{6 \cos^2{\theta} q^2 \left(K_0 z^2\!+\! \tilde{s}\right)}{25 \pi ^2 \Ro_{\mathrm{c}}^2}=0,\\
    &\tilde{s}\left(1-4z^3\right)+3V_0 z^2\left(1-2z^3\right)+\frac{6\cos^2{\theta}\left(\tilde{s}+3K_0 z^2\right)}{25\pi^2 \Ro_{\mathrm{c}}^2}\nonumber\\
    &+2z^5\left(\tilde{s}+V_0z^2\right)\left[3\tilde{s}^2+\tilde{s}\left(4K_0+5V_0\right)z^2+6K_0V_0z^4\right]
  \end{align}
\end{center}

\noindent So one may solve for $q$ from the former equation, whose solution upon
substitution into the latter equation yields an equation solely for the wavevector $z$:

\vspace{-0.25truein}
\begin{center}
  \begin{align}
    &z^3\! \left(V_0 z^2\!+\! \tilde{s}\right)^2\! \left[3V_0 K_0 z^4\!\left(2 z^3\!-\! 3\right)\right.\nonumber\\
    &\qquad\qquad\left.+\tilde{s} z^2\!\left(V_0\!+\! K_0\right)\!\left(4z^3\!-\! 7\right)\!+\! \tilde{s}^2\left(2 z^3\!-\! 5\right)\right]\nonumber\\
    &-\frac{6 \cos^2{\!\theta}}{25 \pi^2 \Ro_{\mathrm{c}}^2}\!\left[2\tilde{s}\left(K_0-V_0\right)+3\tilde{s}^2z\right.\nonumber\\
    &\qquad\qquad\left.+\tilde{s}\left(K_0+5V_0\right)z^3+3K_0V_0z^5\right]\!=\!0.\label{eqn:zeqndiff}
  \end{align}
\end{center}

\noindent The horizontal wavevector is then recovered from the roots of the fourteenth-order Equation
\ref{eqn:zeqndiff}, whereas the superadiabaticity is defined as

\vspace{-0.25truein}
\begin{center}
  \begin{align}
    &\frac{\epsilon}{\epsilon_0} \!=\! \frac{\left(\tilde{s}\!+\!K_0 z^2\right)\! \left(25 \pi^2 \Ro_{\mathrm{c}}^2
      \tilde{s}^2z^5\!\left(\tilde{s}\!+\! V_0 z^2\right)^2\!+\! 6 \cos^2{\theta}\right)}{25\pi^2 \Ro_{\mathrm{c}}^2 \tilde{s}\left(z^3\!-\!
      1\right)\left(\tilde{s}\!+\! V_0 z^2\right)}. \label{eqn:xeqnfd}
  \end{align}
\end{center}

\begin{figure}[t!]
  \begin{center}
    \hspace*{-0.5cm}\includegraphics[width=0.5\textwidth]{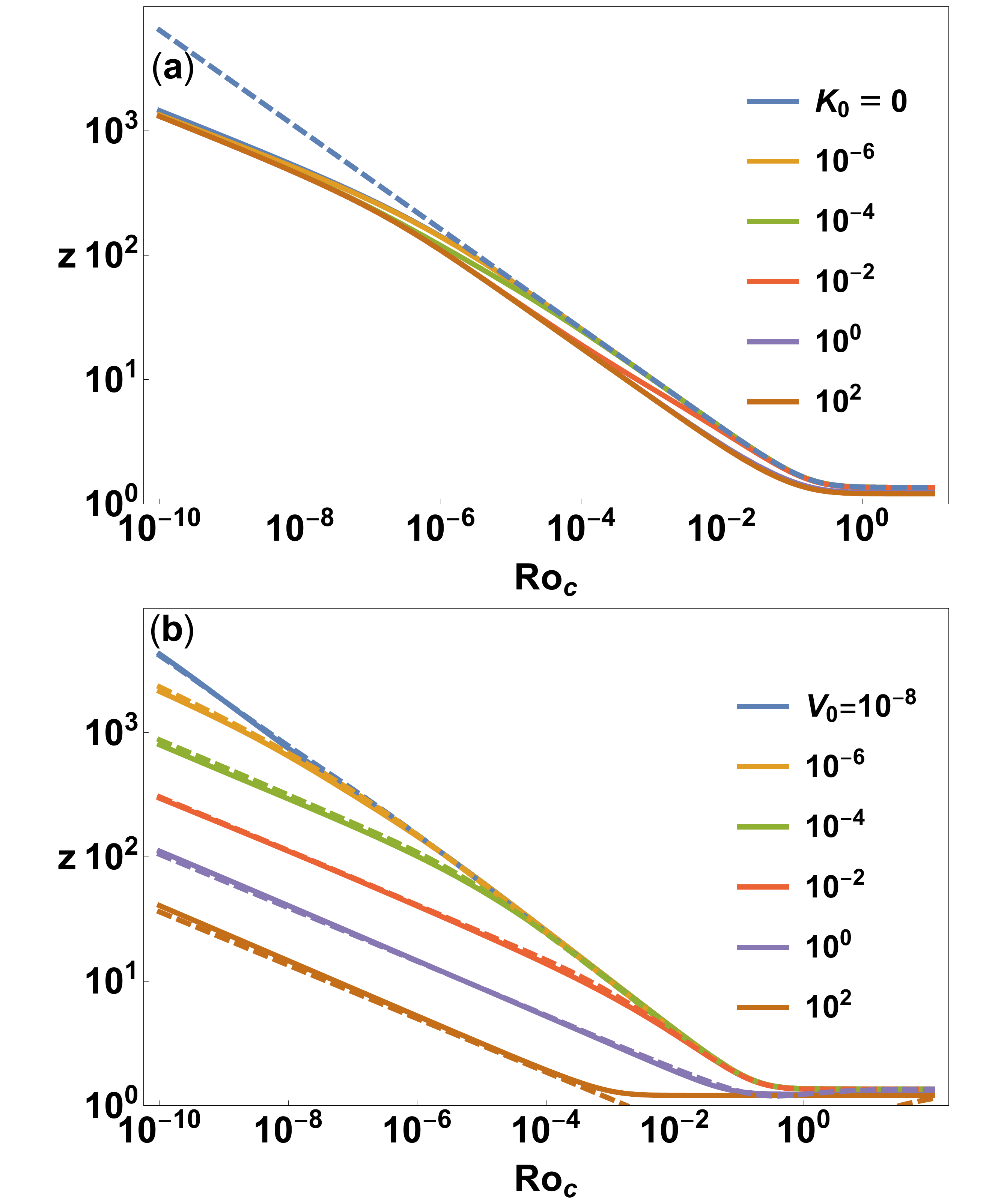} 
    \caption{Convective Rossby number and diffusion dependence of the horizontal wavevector $z$ at the pole
      ($\theta=0$). (a) Scaling of $z$ with convective Rossby number $\Ro_{\mathrm{c}}$ and thermal diffusivity $K_0$, with $V_0=10^{-5}$. The
      dashed line shows the nondiffusive scaling with $\Ro_{\mathrm{c}}$. (b) Scaling of $z$ with viscosity $V_0$ and $\Ro_{\mathrm{c}}$, with
      $K_0=10^{-5}$. The dashed lines show the scaling of the fit given in Equation \ref{eqn:zfit}.}\label{fig:zscaling}
  \end{center}
\end{figure}

Clearly, the inclusion of viscosity increases the degree of superadiabaticity beyond that due to the thermal
diffusion. Moreover, as before, it also grows with lower convective Rossby numbers.  Given the form of Equation \ref{eqn:zeqndiff},
the wavevector must increase with decreasing convective Rossby number, whereas again its behavior with varying $K_0$ is not
obvious, but is relatively benign as is shown in Figure \ref{fig:zscaling}(a). In contrast, viscosity has a much larger
impact on the amplitude of the change in the wavevector as exhibited in Figure \ref{fig:zscaling}(b). The velocity
decreases with lower convective Rossby number given that the wavevector increases, with the same nontrivial behavior seen earlier
in the case with thermal diffusion.  

However, for values of $\left\{K_0, V_0\right\}<N_{*,0}$, one may find a heuristic fit to the horizontal wavevector for the diffusive case
given by

\vspace{-0.25truein}
\begin{center}
  \begin{align}
    &z \!\approx\! 2^{-\frac{11}{6}} 5^{\frac{1}{3}}
      \coth{\!\!\left[3\Ro_{\mathrm{c}}^{\frac{2\sqrt{2}}{5}}\right]^{\frac{\sqrt{2}}{2}}}\tanh{\!\!\left[\frac{17}{10}\Ro_{\mathrm{c}}^{\frac{23}{124}} V_0^{-\frac{37}{160}}\right]}\nonumber\\
    &\left[2+\left(\sqrt{2}\!-\!1\right)\left(\!1\!+\!\tanh{\!\left[\!\frac{\sqrt{2}}{2}\!\left(\frac{3}{2}\!+\!\log_{10}\frac{\Ro_{\mathrm{c}}}{\left(V_0\!+\!
      K_0\right)^{\frac{7}{6}}}\!\right)\!\right]}\!\right)\!\right]\!\!,\label{eqn:zfit}
  \end{align}
\end{center}

\noindent which can be used more directly to estimate the scaling of the relative superadiabaticity and velocity
amplitudes. The form of the fit follows from the observation of the qualitative features of the scaling of the
horizontal wavevector given successively in Figures \ref{fig:steve_scaling}, \ref{fig:diffusive_scaling}, and
\ref{fig:zscaling}, where it is well-described by hyperbolic tangents to within a few percent even in the transitional
region of $\Ro_{\mathrm{c}}\approx 1/10$. The accuracy of the fit is illustrated in Figure \ref{fig:zscaling}(b).

\subsection{Diffusion Approximation for Turbulent Transport}

In stellar models, a diffusive treatment of transport processes can be adopted for thermodynamics, chemicals, angular
momentum, and magnetic field \citep[e.g.,][]{heger00a,maeder09,mathis13a}. Within a convection zone, the turbulent
mixing of those quantities can be approximated through a parameterized vertical diffusion
$D_\mathrm{v} = 1/3 \, \alpha H_P \mathrm{v}_{\mathrm{MLT}}$. Since this depends upon the pressure scale height $H_P$,
it is worth noting that it impacts the superadiabaticity as well. The definition of the Brunt-V\"{a}is\"{a}l\"{a}
frequency is $N^2=g \alpha_T T \epsilon/H_P$. Thus, allowing $H_P$ to vary, one has that
$N^2/N_0^2 = x H_{P,0}/H_{P} = x/h$. To maintain the constancy of the heat flux, this then requires
$x=\epsilon/\epsilon_0$ to depend linearly on $h$, with $x\rightarrow xh$.  

Within the context of this convection model and for a constant mixing-length parameter $\alpha$, the diffusion
coefficient ratio scales as

\vspace{-0.25truein}
\begin{center}
  \begin{align}
	\frac{D_\rmv}{D_0}= h\frac{\rmv}{\rmv_0} = \left(\frac{5}{2}\right)^{\frac{1}{6}}\frac{h}{\sqrt{z}},
  \end{align}
\end{center}

\noindent where $D_0$ is the local diffusion coefficient in the absence of rotation. The scaling of $D_\rmv$ is shown in
Figure \ref{fig:diffusion} and described in detail in the following section, as it suffices to note that the diffusion
will increase with scale-height ratio $h$ and decrease as the rotation rate is increased.  This convection model does
however exclude any horizontal diffusion associated with the increased horizontal shear typically present in rotating
convection.

\section{Convective Penetration}\label{sec:penetration}

\subsection{Models Excluding Rotation}

Standard MLT is not able to address convective penetration directly, rather it must be modified to account for this
nonlocal effect.  Such nonlocalities arise naturally in the generalizations of MLT as they directly model nonlinear
interactions. One of the first and simplest such nonlocal generalizations of MLT can be found in
\citet{spiegel63}. There the convective heat flux is treated as an unknown distribution in position and velocity space
and as such it follows the dynamics described by a collisional Boltzmann equation.  The collisional processes are a
stochastic source arising from nonlinear processes and a damping term.  The latter is the connection to MLT as it limits
the path length of a convective eddy to the mixing length $\ell_{\mathrm{MLT}}$, but it is not required to be small in
some sense as it needs to be in standard mixing length theory.  Another approach is to include the kinetic energy flux,
which leads to a convection theory that depends upon the entropy profile in the star and thus upon global integral
constraints upon the energy flux in the star \citep[e.g.,][]{roxburgh78}. Other models have also been developed that
variously extend MLT to include nonlinear processes through finite amplitude analyses
\citep[e.g.,][]{veronis63,sparrow64}. Subsequently, modal expansions of the Boussinesq and anelastic equations provided
solutions that established that nonlinear penetration was substantially deeper than what linear theory predicted
\citep[e.g.,][]{musman68,moore73}. In anelastic convection, the density stratification induces asymmetries between
upflows and downflows that further increases the penetration depth \citep[e.g.,][]{zahn82,massaguer84}.  Yet as is often
the case, the penetration depth determined in these models depends sensitively on their assumptions, with a wide
disparity in the calculated penetration depths. The problems and inconsistencies with many of these models were reviewed
in \citet{renzini87}.

Three pioneering studies analytically estimate the convective penetration depth by prescribing the convective flows as a
boundary condition and examining the impact on the region of penetration. One of the first models to attempt to estimate
the importance of overshooting above a convective core is from \citet{roxburgh65}, in which ergodicity and isotropy are
assumed, leaving a monodimensional and time-independent equation of motion. A linearization of the thermodynamics with
respect to the vertical displacement then permits the equation of motion to be integrated to yield an estimate of the
depth of penetration of a fluid element. Building upon the work of \citet{roxburgh78}, \citet{schmitt84} uses analytical
models of turbulent plume ensembles and examines their statistical properties, finding that the penetration depth scales
as $\rmv_p^{3/2} f^{1/2}$, where $\rmv_p$ is the plume velocity and $f$ is the fractional area filled by the
plumes. Likewise, following \citet{roxburgh65} and \citet{roxburgh78} and by directly parameterizing the filling factor
$f$, \citet[][hereafter Z91]{zahn91} found a similar scaling result through a spatial linearization of the equations of
motion and the energy transport. Both of these studies predict penetration depths as fairly large fractions of a
pressure scale height. \citetalias{zahn91} also confirmed the modal results of \citet{massaguer84} where the flow
asymmetry due to the density stratification causes downflows to penetrate farther into an underlying stable layer than
upflows penetrate into an overlying layer. As a first estimate of the convective penetration depth, one may consider the
\citetalias{zahn91} linearized model for convective penetration. This model neglects the effects of rotation in both the
physical description of the model as well as in its parameterization of the convective dynamics.  The former
modification will be saved for later work. However, the extended version of the \citetalias{stevenson79} model derived
above provides a means of estimating the latter alteration. In particular, one can harness the linearized
framework of the \citetalias{zahn91} model to obtain an order of magnitude estimate of the depth of penetration while
allowing for the effects of rotation to be included through a modified value of the superadiabaticity and mixing length
velocity within the convective region. It is also a formalism that may be easily implemented in stellar evolution
codes.

\subsection{Convective Penetration with Rotation}

\subsubsection{Theory}

The \citetalias{zahn91} model has four phenomenologically distinct zones: a convective zone that ends when the radiative
energy flux becomes equal to the convective energy flux, a subadiabatic penetrative zone where flows render the region
nearly adiabatic, a transitional thermal boundary layer with overshooting convection where the P\'{e}clet number is
reduced to below unity, and a radiative zone where waves and conduction carry the total energy flux. There are a few
basic assumptions underlying the model.  The first assumption is that the velocity and the temperature share the same
planform, namely that they are perfectly correlated. The second assumption is that only the downflows for downward
penetrating flows and upflows for upward penetrating flows are effective at carrying enthalpy, which is parameterized
through their filling factor $f$, which results from a horizontal average over their planform
$\vcz(x,y,z) = \rmv(z) h(x,y)$ such that $f=\overline{h^2(x,y)}$.  From simulations, the value of $f$ appears to be
about $1/3$, with slight variations of order of 10\% that are sensitive to the level of turbulence, rotational, and
magnetic effects \citep{brummell02,stein09a,stein09b}.  However, rotation does impact the structure of the convection as
a function of colatitude, whose structure in turn depends upon the degree of supercriticality of the system.  For
rotating systems, the convective structures should possess a modicum of alignment with the rotational axis, which
implies that the mechanism of penetration will vary with colatitude.  For instance, consider that columnar flow
structures will mostly conform to the classical paradigm of radially-aligned penetrating flows near the pole, but they
will increasingly become horizontally-shearing, yet still localized, flows at lower colatitudes.  This will be combined
with any shear-induced mixing by large-scale flows such as differential rotation
\citep[e.g.,][]{zahn92,maeder03,mathis04b,mathis13a}. Yet, for the purposes of this model, the filling factor $f$ shall
be considered to be a fixed parameter of the theory. Additionally, it shall be assumed that the local magnitude of the
gravity $g$, the local conductivity $\kappa$, the thermal expansion coefficient $\alpha_T$, the nuclear energy
generation rate $Q$, and the specific heat capacity at constant pressure $c_P$ are unaffected by rotation, namely that
the centripetal acceleration may be ignored to first order.

A further assumption is that the convective enthalpy flux can be linearized in the region of penetration such that

\vspace{-0.25truein}
\begin{center}
  \begin{align}
    F_{\mathrm{conv} } &= \rho c_P \overline{\vcz T'} = f\rho c_P \rmv \delta T \label{eqn:fconv}\\
    &= F_0\left[\frac{d\ln{M}}{d\ln{r}}-\frac{d\ln{L}}{d\ln{r}}+\frac{d\ln{\chi}}{d\ln{r}}\right]_{r_0} \frac{z}{r_0},
  \end{align}
\end{center}

\noindent where $M(r)$ is the mass, $L(r)$ is the luminosity, and $\chi(r)$ is the radiative conductivity such that
$\chi = \rho c_P \kappa$ as seen in \citetalias{zahn91} (Equation 4.4).  Finally, another assumption is that the vertical
inertial force balances the buoyancy force on average such that the pressure contribution vanishes and so that the
density perturbations can be linearized with $\rho' = \alpha_T T'$. To follow \citetalias{zahn91} as closely as
possible, the system is considered only at the pole so that the direct effects of the local Coriolis acceleration
$2\Omega_0\sin{\theta}v_x$ may be neglected, which is equivalent to having a buoyant braking timescale that is shorter
than the rotational timescale. Instead, the Coriolis effect implicitly influences the penetration depth by modifying the
upper boundary value of the velocity, which is taken from the convection zone. Thus, when horizontally-averaged and a
pressure equilibrium is assumed as in \citetalias{zahn91} (Equation 3.8), the force balance becomes

\vspace{-0.25truein}
\begin{center}
  \begin{align}
    \frac{c}{2}\frac{d\rmv^2}{dz} = g \frac{\delta T}{T},
  \end{align}
\end{center}

\noindent where $c=\overline{h^3}/\overline{h^2}$.  This equation can then be linked to the convective energy flux
through Equation \ref{eqn:fconv} and integrated across the penetrative region to yield an estimate for the penetration
depth $L_P$ relative to the pressure scale height $H_P$.

In \citetalias{zahn91}, there are two cases for the scaling of the convective penetration depth with velocity and
spherically-symmetric thermodynamic quantities: one for penetrative convection into a stable region below a convection
zone, such as takes place near the base of the solar convection zone, and one for penetrative convection into a stable
region above the convection zone, such as takes place near the convective cores of intermediate mass and high mass
stars.  The first of these two regimes neglects the variation in the total luminosity and mass and so only retains the
term related to changes in the radiative conductivity. It scales as

\vspace{-0.25truein}
\begin{center}
  \begin{align}
    &\frac{L_P}{H_P} \!=\! \left[\frac{2}{3}\frac{\left(1-f\right) f \rmv_z^{3}}{g \alpha_T \kappa\chi_P\nabla_{\mathrm{ad}}}\right]^{\frac{1}{2}},
  \end{align}
\end{center}

\noindent where $\nabla_{\mathrm{ad}}$ is the adiabatic temperature gradient and
$\chi_P=\partial \ln{\kappa}/\partial \ln{P}|_{S}$ is the adiabatic logarithmic derivative of the radiative conductivity
with respect to pressure.  The opposite case of convective penetration into a stable region above a convection zone
scales as

\vspace{-0.25truein}
\begin{center}
  \begin{align}
    &\frac{L_P}{H_P} \!=\! \left[\frac{2\left(1-f\right) f \rmv_z^{3} r_0}{9g \alpha_T \kappa
    \nabla_{\mathrm{ad}} H_P\left(\rho P/\overline{\rho P} - \rho Q/\overline{\rho Q}-r_0\chi_P/3H_P\right)}\right]^{\frac{1}{2}}\!\!\!,
  \end{align}
\end{center}

\noindent where $r_0$ is the radius at the edge of the core. Note that $\nabla_{\mathrm{ad}} = \nabla + \epsilon$, with
$\nabla$ being the temperature gradient and $\epsilon$ being the superadiabatic gradient as above.  However, the basic
assumption of the model is that $\epsilon$ does not grow large enough to modify the background temperature gradient in a
steady state, and so its variation with rotation rate does not strongly influence the depth of penetration. Thus, the
ratio of the penetration depth with rotation and diffusion to the nonrotating inviscid value for convective penetration
into a stable layer either above or below a convection therefore scales as

\vspace{-0.25truein}
\begin{center}
  \begin{align}
    &\frac{L_P}{L_{P,0}} \!=\! \left(\frac{\rmv}{\rmv_0}\right)^{3/2} = \left(\frac{5}{2}\right)^{\frac{1}{4}}z^{-\frac{3}{4}}.
  \end{align}
\end{center}

\begin{figure}[t!]
  \begin{center}
    \hspace*{-0.5cm}\includegraphics[width=0.5\textwidth]{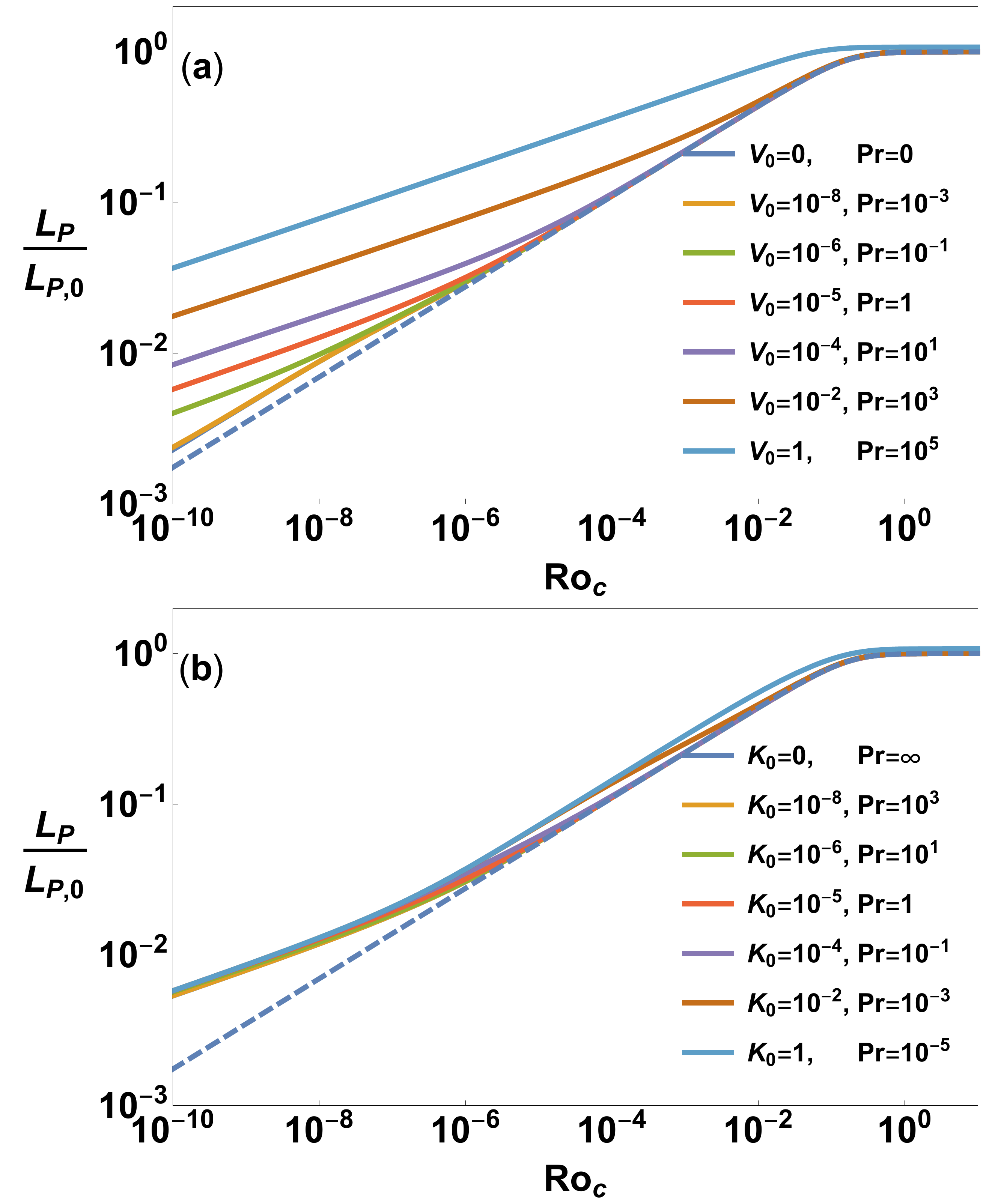} 
    \caption{Convective Rossby and Prandtl ($Pr=V_0/K_0$) number dependence of the convective penetration depth $L_P$ at
      the pole ($\theta=0$). (a) Scaling of $L_P$ with viscosity $V_0$ at a fixed thermal diffusivity $K_0=10^{-5}$.
      (b) Scaling of $L_P$ with $K_0$ with $V_0=10^{-5}$.}\label{fig:lp_scaling}
  \end{center}
\end{figure}

As seen in the previous section, the velocity amplitude of the mode that maximizes the heat flux decreases with lower
diffusivities and lower convective Rossby numbers. Therefore, the penetration depth necessarily must decrease when the convective Rossby
number is decreased. This behavior follows intuitively given that the reduced vertical momentum of the flows implies
that the temperature perturbations are also reduced (via Equation \ref{eqn:v_theta_link}). Thus, due to the decreased
buoyant thermal equilibration time and the reduced inertia of the flow the penetration depth must decrease. In contrast,
the velocity and the horizontal scale of the flow increase with greater diffusivities in order to offset the reduced
temperature perturbations in the case of a larger thermal conductivity.  In the case of a larger viscosity, the
horizontal scale of the velocity field is increased, whereas, for a fixed thermal conductivity, the thermal
perturbations are of a smaller scale. Thus, to maintain the heat flux, the amplitude of the velocity must increase in
order to compensate for the reduced correlations between the two fields. The scaling behaviors of the penetration depth
are illustrated as a function of diffusivities and convective Rossby number in Figure \ref{fig:lp_scaling}.

\subsubsection{Comparison with Penetrative Convection Experiments}\label{sec:compare}

Penetrative convection has been studied extensively in fully compressible and anelastic numerical experiments in 2D
within the context of the solar convection zone \citep{hurlburt86,hurlburt94,rogers05a,rogers06}, for the convective
cores of intermediate mass stars \citep{rogers13}, and in the context of modeling laboratory experiments for
temperature stratified water \citep{lecoanet15,couston17,toppaladoddi17}. These studies all tend to
find that penetration into stable layers below the convection zone is more extensive than into overlying stable layers
and that the depth of penetration depends sensitively on the stiffness of the interface between them and upon the
strength of the convective driving.  Moreover, in rotating convection models of equatorial planes of intermediate mass
stars, it is directly seen that rotation reduces the overshooting depth as the Coriolis force leads to an azimuthal
deflection of convective plumes and to less available radial kinetic energy \citep{rogers13}. While 2D simulations are a
useful first step to understanding convective penetration, they are limited in assessing the convective properties of
stellar interiors given the 3D nature of the convection there, which leads to much different convective structures and
spectral energy transfer properties.

In constrast to the behavior of penetrative convection seen in local simulations and from scaling arguments such as
those described above, the depth of penetration in fully nonlinear global-scale 3D simulations tends to increase with
decreasing convective Rossby number and in spherical geometry
\citep[e.g.,][]{miesch00,miesch05,browning04,brun11,augustson12,augustson13,alvan14,augustson16,brun17a}.  However,
these simulations are typically in a low P\'{e}clet number regime $Pe \approx\mathcal{O}(10)$ and thus are still
influenced by thermal diffusion. The stiffness of the interface often has been a priori softened so that the simulation
is more computationally feasible, meaning that the degree of convective overshoot relative to penetration is larger than
in a stellar context and that the excitation mechanisms are also more dominated by Reynolds stresses rather than
buoyantly driven.  So, it may be that the stiffness of the transition region is too weak in current global-scale
modelling approaches, permitting large-scale entraining and overshooting flows rather than restricting them to
plume-like dynamics.  Moreover, the sweeping motions of the relatively laminar globally connected flows induce mixing on
a scale of the correlation length of those flows \citep{viallet15}, which enhances the depth of penetration and is
something that will be assessed in later work.

Three-dimensional simulations have been conducted that can simultaneously resolve gravity waves as well as the
mechanisms that excite them, namely shear and convective penetration in both nonrotating Cartesian settings
\citep[e.g.,][]{hurlburt94,lecoanet15,couston18} and f-planes \citep[e.g.,][]{julien96,brummell02,pal07,pal08}.  As in
2D simulations, these 3D simulations also find the following: penetration into stable layers below the convection zone
is more extensive than into overlying stable layers, the depth of penetration depends sensitively on the stiffness of
the interface between them, and the penetration depth is latitudinally dependent.

In the 3D f-plane simulations of rotating convection described in \citet{julien96} and \citet{brummell02}, it is found
that the penetration depth into a stable layer below a convective region scales as
$L_P \propto \Ro_{\mathrm{c}}^{0.15}$, that is it decreases with increasing rotational influence, due primarily to a
reduction in the flow amplitude.  Likewise, in a similar suite of f-plane simulations examined in \citet{pal07}, it is
found that there is a decrease in the penetration depth with increasing rotation rate that scales as
$L_P \propto \Ro_{\mathrm{c}}^{0.2}$ at the pole and to $L_P \propto \Ro_{\mathrm{c}}^{0.4}$ at mid-latitude.  For
penetration into a stable layer above a convection zone, on the other hand, f-plane simulations from \citet{pal08}
indicate that there is a much weaker rotational scaling, being statistically consistent with no scaling. However, the
range of parameters examined is quite restricted due to the computational requirements to resolve structures as well as
maintain highly supercritical flows at much lower convective Rossby numbers. Nevertheless, the depth of convective
penetration as assessed in those numerical simulations appears to be roughly consistent with the heuristic model derived
above, where $L_P/L_{P,0} \propto \Ro_{\mathrm{c}}^{3/10}$, which follows from
$\rmv/\rmv_0 \propto \Ro_{\mathrm{c}}^{1/5}$ in the nondiffusive and low convective Rossby number limit of the
convection model.

\citet{couston17} have examined the influence of the stiffness of the convective-radiative transition in numerical
analogs of a laboratory experiments involving the buoyancy transition of water that occurs near $4^\circ$~C, which is
also a decent analog for convective stellar cores.  In these 2D simulations, there is a smooth transition between plume
dominated and entrainment dominated dynamical regimes that is independent of the Peclet number.  Instead, it is
sensitive to the stiffness of the convective-radiative transition region and the Rayleigh number.  According to those
local domain numerical simulations, and to the theory developed in \citet{hurlburt94}, the interface stiffness provides
a further distinction drawn between the regimes of penetrative convection. Particularly, the penetration depth depends
upon the relative stability $S=(2 n_s-3)/(3-2 n_c)$ between the stably-stratified and convective regions, where $n_s$ is
the polytropic index in the stable region and $n_c$ is the index in the convective region.  Indeed, \citet{hurlburt94}
employs a linearized convection model similar to the one derived in \citetalias{zahn91}, which has a continuous thermal
conductivity, but instead the conductivity is taken to be piecewise constant. Using this model, it can be shown that the
total penetration depth depends upon the depth of the adiabatic layer and the depth of the thermal adjustment layer.
However, the depth of the two regions scale differently with the relative stability parameter $S$.  In particular, the
depth of the adiabatic layer scales as $L_a \propto S^{-1}$ and the thermal adjustment layer scales as
$L_t \propto S^{-1/4}$.  Therefore, as $S$ increases the depth of the adiabatic layer decreases significantly, leaving
the thermal adjustment layer as the primary means of defining the depth of penetration.  Note that as stated in
\citet{zahn91} (Equation 3.18), the penetration model developed there and expanded on above scales proportionally to the
inverse of the interface stiffness. For the simulations of penetration into a stably-stratified layer below a convection
zone, \citet{brummell02} and \citet{pal07} find that the depth of penetration scales primarily as
$L_P \propto S^{-1/4}$, for values of $S$ between about unity and around 30.  Yet for simulations of convective
penetration into a stably-stratified region above a convection zone, the opposite behavior is found, namely that
$L_P \propto S^{-1}$ for $S$ between unity and ten. Given the similarity in the parameters used in those simulations,
one might conclude that the penetration into a stably-stratified layer below a convection zone has physics more akin to
a thermal adjustment layer, whereas penetration into a layer above is more akin to an adiabatic penetrative layer.

\begin{figure}[t!]
  \begin{center}
    \vspace*{1cm}\includegraphics[width=0.45\textwidth]{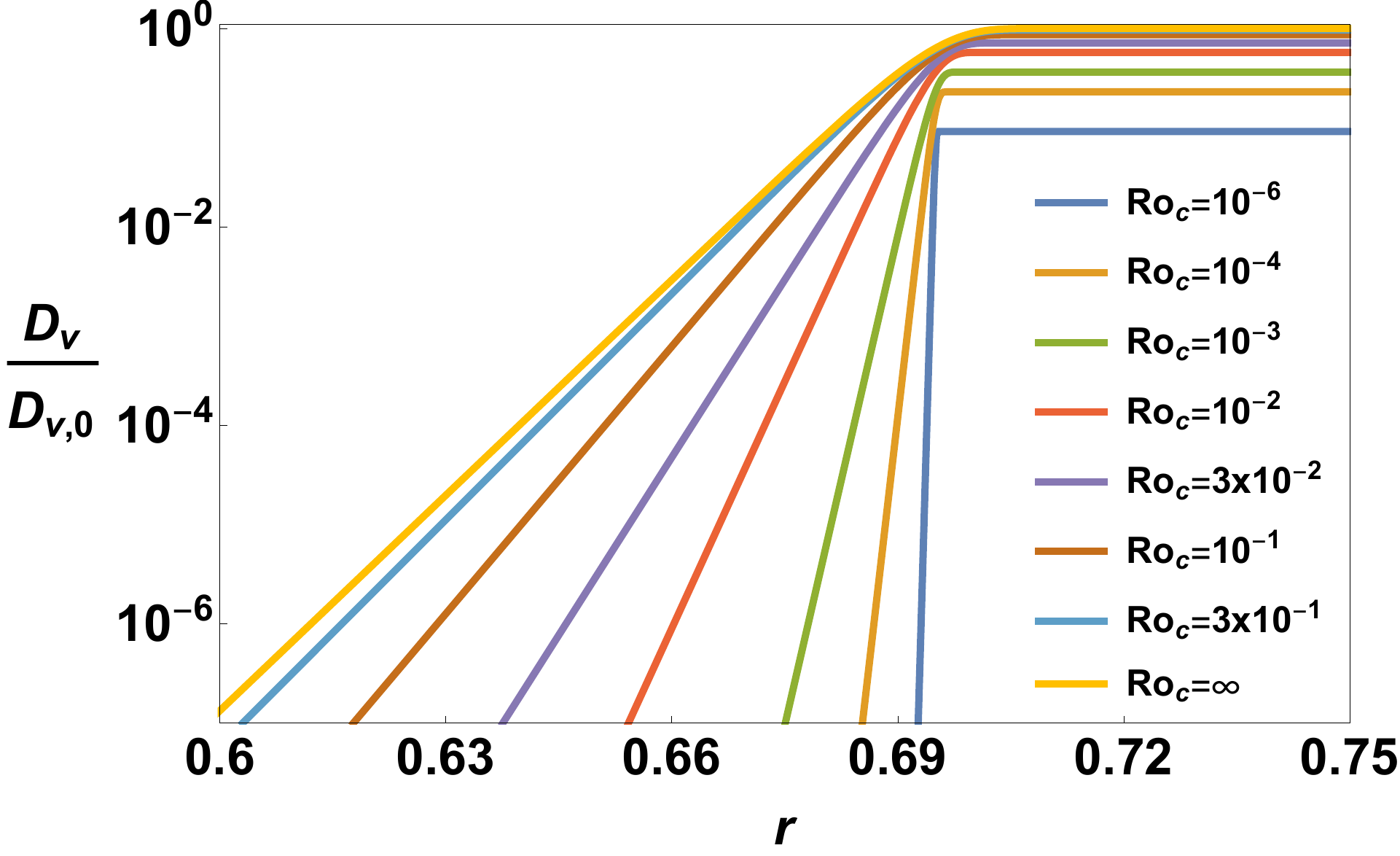} 
    \caption{The radial dependence of the vertical mixing length diffusion coefficient for a solar-like star near the
      transition between the convectively stable and unstable layers for the inviscid convection model, showing the dual
      effects of decreased diffusion with decreasing convective Rossby number and the increasing lower radial limit of
      the diffusion coefficient due to the decreasing depth of penetration.}\label{fig:diffusion}
  \end{center}
\end{figure}

\subsection{A Diffusive Approach}

This model of penetrative convection can be introduced into stellar models utilizing a diffusive approach.  Such a
parameterization of mixing processes has been extensively examined
\citep[e.g.,][]{freytag96,herwig99,denissenkov13,viallet15,lecoanet16}. A complementary model has been established
through an extreme-value statistical analysis of 3D penetrative convection simulations \citep{pratt17a}.  A frequent
assumption made of turbulent flows is to assume that the underlying distribution of a flow quantity is normal. However,
it is shown through the simulations and analysis of \citet{pratt17a} and \citet{pratt17b} that the more rare events
occurring in the tails of the actual distribution function, e.g. the extremal penetrative flows with a higher velocity
and a greater entropy deficit than the average flow, have a larger impact than is expected in Gaussian turbulence models
on the turbulent mixing coefficients.  This analysis was carried out with a novel method to quantify this non-Gaussianity
of the dynamics. Their method permits a statistical examination of the turbulent particle dispersion and subsequently
the construction of a model for a turbulent diffusion based upon the Gumbel distribution \citep{pratt17b}. This model
for diffusion and overshoot has been applied to the problem of lithium depletion occurring in rotating low-mass stars
\citep{baraffe17}, where it was found to mimic the observed trends well when the effects of rotation on the diffusion
treatment were introduced.  There the rotational effects are included as an empirically defined threshold rotation rate
above which the diffusion is quenched ostensibly due to the action of rotation on the convective flows.  Also, the
penetrative convective depth was a parameter and was not dynamically estimated.  However, the initial estimate of
\citet{baraffe17} may be improved. Using the above extension of the \citetalias{zahn91} model, one can estimate both the
penetration depth and the level of diffusion. Taking the parameters of the Gumbel distribution as in \citet{pratt17b},
yields the following description of the radial dependence of the diffusion coefficient

\vspace{-0.25truein}
\begin{center}
  \begin{align}
  	&D_\rmv\left(r\right) =\left(\frac{5}{2}\right)^{\frac{1}{6}}\!\frac{\alpha H_{P} \rmv_0 h}{3\sqrt{z}} \nonumber\\
    &\qquad\left\{1\!-\!\exp{\!\left[-\exp{\left(\!\left(r-r_c\right)/\lambda L_P\!+\!\mu/\lambda\right)}\right]} \right\},\label{eqn:pdiff}
  \end{align}
\end{center}

\noindent where $r_c$ is the base of the convection zone and where $\mu$ and $\lambda$ are the empirically determined
parameters from \citet{baraffe17}.  An illustration of the scaling behavior due both to the variation of $D_\rmv$ with
the mixing length velocity $\rmv$ and with the depth of penetration $L_P$ are shown in Figure \ref{fig:diffusion}, where
the parameters are taken to make contact with Sun-like stars where the transition region begins around
$r\approx 0.7 R_{\sun}$. Note, however, the convective velocity will vary in the convection zone. So Figure
\ref{fig:diffusion} is only meant to illustrate the rough behaviors of this diffusive model near the convectively
stable-to-unstable transition point. From Equation \ref{eqn:pdiff} and as seen in Figure \ref{fig:diffusion}, the radial
structure of the diffusion coefficient follows from the scaling of the velocity, namely the diffusion will globally
decrease with decreasing convective Rossby number, and will further vary if the local angular velocity is taken into
account.  The depth of penetration is perhaps most notable, in that its strong rotational dependence can lead to severe
restrictions on the region in which the diffusion acts. This is evidenced through the transition between a long
diffusive tail at high convective Rossby number to a step-like function at low convective Rossby number.  Thus, as
discussed earlier, the amount of and the depth over which mixing induced above or below a convective region may be
severely reduced in stars whose convection possesses a low convective Rossby number.

In \citet{lecoanet16}, the mixing induced below a convection zone is characterized as a means of determining whether or
not carbon flames can be disrupted by overshooting convection in evolved massive stars of between about 7-11 solar
masses. To estimate the amount of mixing, the results of an idealized model of these flames are assessed with 3D
hydrodynamic simulations of the Boussinesq equations. In particular, a passive scalar field is evolved that permits the
measurement and fitting of a height-dependent turbulent chemical diffusivity that mimics the effect of the overshooting
convection.  This diffusion coefficient is modeled as a composition of two error functions, which has a
super-exponential tail in the mixing region somewhat similar to the model constructed above.

What is currently lacking, however, are extensive parameter studies using 3D simulations of convective overshoot that
also include rotation. Such simulations would at least provide numerical evidence regarding the hypothesized dependence
of the mixing depth and its shape. For instance, the simulations carried out in \citet{korre18} have also addressed the
shape of the overshooting region in a Boussinesq system and its associated potential mixing properties in 3D
simulations.  As in \citet{lecoanet16}, it is found that the horizontally-averaged kinetic energy is super-exponentially
reduced with depth below the convection zone. Specifically, in those simulations, the average kinetic energy is best
modelled with a Gaussian. Moreover, the radial correlation length scale downflows is larger than the length scale
associated with the averaged kinetic energy profile. This length scale may be interpreted as depth to which the
strongest downflows travel before stopping. This behavior is at least roughly consistent with the model proposed here,
and with the results of \citet{pratt17a} and \citet{pratt17b}, although it is in contrast to the models proposed in
\citet{freytag96} and \citet{herwig00}. For the time being, the examinations in \citet{julien96} and \citet{brummell02}
will have to suffice as evidence that the overshooting is reduced with rotation rate, while the shape of the mixing
region likely depend upon the model setup and equations solved in similar simulations.

\section{Summary and Discussion}\label{sec:final}

A model of rotating convection originating with \citet{stevenson79} has been extended to include thermal and viscous
diffusion for any convective Rossby number.  Moreover, a systematic means of developing such models for an arbitrary
dispersion relationship have also been shown.  An explicit expression is given for the scaling of the horizontal wavenumber
in terms of the convective Rossby number and diffusion coefficients under the constraint that the values of the
diffusive time scale are less than the convective time scale (Equations \ref{eqn:zsteve}, \ref{eqn:zthermal}, and
\ref{eqn:zeqndiff}).  The scalings of the velocity and superadiabaticity in terms of that wavenumber are also given.
Asymptotically at low convective Rossby number and without diffusion, these match the expressions given in
\citet{stevenson79} (see Figure \ref{fig:steve_scaling}), as well as the numerical results found in the 3D simulations
of \citet{kapyla05} and \citet{barker14}.

In \S\ref{sec:penetration}, the model of rotating convection is employed to assess the convective Rossby number scaling
of the depth of convective penetration, utilizing the linearized model of \citet{zahn91}. Due to the reduced velocity
and increased superadiabaticity at lower convective Rossby number, the penetration depth decreases proportionally to
$\Ro_{\mathrm{c}}^{3/10}$ when diffusive processes are neglected. This estimate of the penetration depth is then
employed to construct an estimate for the diffusive mixing coefficient in the region of penetration based upon the
numerical experiments of \citet{pratt17a} and \citet{pratt17b}. In the 3D f-plane simulations of rotating convection of
\citet{julien96} and \citet{brummell02} it is found that $L_P\propto\Ro_{\mathrm{c}}^{0.15}$. Similarly, in the
simulations of \citet{pal07} and \citet{pal08} it is found that $L_P$ varies in latitude, with
$L_P\propto\Ro_{\mathrm{c}}^{0.2-0.4}$ from the pole to mid-latitudes. Hence, the depth of convective penetration as
assessed in those numerical simulations is roughly consistent with those that follow from the coupling of the convection
model with the results of \citet{zahn91}.

This convection model will soon be exploited to begin to study the impact of reduced convective penetration in stellar
and planetary evolution models when rotation is included. Yet its impact on the structure of the convection can already
be anticipated. From Figure (\ref{fig:diffusive_scaling}), since this convection model is local, a qualitative picture
of those aforementioned impacts can be constructed as follows: in regions where the local convective Rossby number is
less than unity, the mixing-length velocity will be reduced and the superadiabatic temperature gradient will be
enhanced.  Therefore, depending upon the value of the superadiabatic gradient in the nonrotating case, the full
temperature gradient may increase sufficiently to modify the location of the ionization zones as well as regions of
large opacity changes.  However, most of those thermodynamic changes will be felt most keenly deeper within the
convection zone where the convective Rossby number can be smaller than near the photosphere for stars, where the
convective Rossby number is typically of order unity or larger.  Near the boundary of convective regions the typical
mixing-length velocity is small due to the increasing importance of the radiative transport of energy.  Thus, in those
regions, the local convective Rossby number can be quite small and yet the thermal diffusivity can become larger
relative to the dynamical time scale.  With that in mind, and again appealing to Figure (\ref{fig:diffusive_scaling}),
it is clear that the velocity will be even further reduced and the superadiabatic gradient further increased. This
implies that the transition to the region of convective penetration and the radiative zone itself will be shallower (or
deeper for a convective core) when compared to the nonrotating case.  Hence, in addition to the reduced depth of
convective penetration with decreasing convective Rossby number seen in \S\ref{sec:penetration}, the convection zone
depth will be reduced when considering convection in the presence of rotation with this model. Due to decreased
diffusion and transport, sharper thermodynamic and chemical gradients may be present in convectively stable regions.

Being a first step, there are several improvements that can be made to this convection model. One clear omission in this
work is the impact of the magnetic field. However, magnetism has been considered in both \citet{stevenson79} and
\citet{canuto86}. An alternative model for the impact of the magnetic field on the superadiabaticity, derived from the
variational principle of \citet{bernstein58}, can be found in \citet{gough66}. Thus, convection and magnetic field will
be the focus of a paper in this series.  Second, the conjecture of \citet{stevenson79} gives the scaling of only the
dominant heat-carrying mode.  To build a theory of the turbulent spectrum and to assess the changing cascade of energy
and helicity, one could extend the theory in a manner similar to that shown in \citet{malkus54}, with a mode by mode
analysis of the heat flux to construct its accompanying spectrum.  One attempt at generalizing this turbulence model has
already been undertaken utilizing the Heisenberg-Kolmogorov turbulence model, as shown in \citet{canuto86}. It would
also be prudent to further numerically assess the validity of these scaling laws with 3D simulations with a larger range
of Nusselt number and to examine the impact of diffusion. The theory can be further improved by considering more
sophisticated models of the structure of the flows, such as applying the results obtained for rotating plumes considered
in \citet{pedley68} or \citet{grooms10}.  Additionally, one relatively simple improvement to the treatment of convective
penetration will be to include the effects of rotation in the linearized equations of motion in an f-plane, expanding
further upon the \citet{zahn91} model.

These waves, which are internal gravity waves modified by rotation through the Coriolis acceleration
\citep[e.g.,][]{dintrans00}, propagate in stably stratified stellar radiation zones. They provide a means for
transporting angular momentum and mixing chemical species in these regions
\citep[e.g.,][]{pantillon07,mathis08,mathis09,lee14}. They are one of the mechanisms that may explain the strong
extraction of angular momentum in the Sun \citep[e.g.,][]{talon05}, in subgiant and red giant stars
\citep[e.g.,][]{fuller14,belkacem15a,belkacem15b,pincon17}, and in intermediate-mass and massive stars
\citep[e.g.,][]{rogers15} as revealed by the weak differential rotation observed thanks to helio- and asteroseismology
\citep[e.g.,][]{garcia07,beck12,mosser12,deheuvels12,kurtz14,saio15,murphy16,aerts17,gehan18}. However, their amplitude
and frequency spectrum strongly depend on the properties of the turbulent convective flows that excite them
stochastically at the radiation/convection interface and in the bulk of convective regions
\citep[e.g.,][]{schatzman93,zahn97,rogers06,belkacem09a,lecoanet13,rogers13,alvan14}. Therefore, one should
consider the impact of the modification of their convective excitation source by rotation. Although some first attempts
have been made to do this \citep{belkacem09b,mathis14,rogers15}, a systematic study should be done. This will be adressed
in the second article of this series.

Another possible use of the convection model developed here is for tidal dissipation. Turbulent friction is one of the
crucial physical mechanisms driving the dissipation of tidal flows in stellar and planetary convective regions
\citep[e.g.,][]{zahn66,zahn89,ogilvie12}. This friction acts both on the equilibrium tide and on tidal inertial waves in
these layers. Recently, in \citet{mathis16}, the asymptotic version of Stevenson's theoretical scaling laws was used to
construct a corresponding local model of tidal waves in order to understand the consequences of rotating convective
turbulence for linear tidal dissipation. With this convective model, the turbulent friction acting on the tides was
significantly reduced in rapidly rotating objects. Thus, to better capture the behavior of the tidal friction in the
convective regions in stars and planets at modest convective Rossby number, it will be useful to apply the model of
convection developed here.

\section*{Acknowledgments} {The authors thank the anonymous referee for their constructive comments, which have improved
  the description of the convection model and its implications. K.~C. Augustson and S. Mathis acknowledge support from
  the ERC SPIRE 647383 grant and PLATO CNES grant at CEA/DAp-AIM.  The authors also thank Q. Andr\'{e}, A. Astoul,
  M. Browning, A.~S. Brun, V. Prat, R. Raynaud, and J. Toomre for fruitful conversations.}

\bibliography{penetrative_convection}

\end{document}